\documentclass[11pt]{article}
\usepackage{latexsym}
\usepackage{amsmath}
\usepackage{amsfonts}
\usepackage{mathrsfs}
\usepackage{dsfont}
\usepackage{enumerate}
\usepackage{verbatim}
\usepackage{hyperref}
\usepackage{tikz}
\usepackage{mathtools}
\usepackage{simplewick}
\usepackage{float}
\usepackage{caption}
\usepackage{slashed}
\usepackage[title]{appendix}
\usepackage{graphicx}
\usepackage{subfig}
 \csname
@addtoreset\endcsname{equation}{section}
\usepackage{amssymb}
\usepackage{color}
\usepackage{wasysym}
\usepackage{array}
\begin{document}

\def\be{\begin{equation}}
\def\ee{\end{equation}}
\def\bea{\begin{eqnarray}}
\def\eea{\end{eqnarray}}
\def\nn{\nonumber}

\renewcommand{\thefootnote}{\fnsymbol{footnote}}
\renewcommand*{\thefootnote}{\fnsymbol{footnote}}

\begin{flushright}

\end{flushright}

\vspace{40pt}

\begin{center}

{\Large\sc Tidal Love numbers of braneworld black holes and wormholes}

\vspace{50pt}

{\sc Hai Siong Tan}

\vspace{15pt}
{\sl\small Division of Physics and Applied Physics,
School of Physical and Mathematical Sciences, \\
Nanyang Technological University,\\
21 Nanyang Link, Singapore 637371\\[5ex]
Physics Division, National Center for Theoretical Sciences,\\
Hsinchu 30013, Taiwan. }

\vspace{15pt}

\vspace{70pt} {\sc\large Abstract}\end{center}

We study the tidal deformations of various known black hole and wormhole solutions in a simple context of warped 
compactification  ---  
Randall-Sundrum theory in which the four-dimensional spacetime geometry
is that of a brane embedded in five-dimensional Anti-de Sitter spacetime. 
The linearized gravitational perturbation theory generically reduces to either
an inhomogeneous second-order ODE or
a homogeneous third-order ODE of which 
indicial roots associated with an expansion about asymptotic infinity can be related to Tidal Love Numbers.
We describe various tidal-deformed metrics, classify their indicial roots,
and find that in particular the quadrupolar TLN is generically non-vanishing. 
Thus it could be a signature of a braneworld by virtue of its potential appearance in gravitational waveforms 
emitted in binary merger events.

\newpage

\tableofcontents

\renewcommand*{\thefootnote}{\arabic{footnote}}
\setcounter{footnote}{0}

\section{Introduction}
\label{Intro}
Recent spectacular advances in gravitational-wave (GW) detection 
have brought new optimism and excitement in using GW astronomy 
to explore a range of topics in black hole and neutron star physics.
In particular, the GW signal emitted by a neutron-star binary carries
information about the nuclear equation of state through Tidal Love Numbers
which are fundamentally a set of quantities that measure the gravitational 
tidal deformation of the object. 
As was first explained in a seminal work by Flanagan and Hinderer in 
\cite{Flanagan}, the (electric, quadrupolar) TLN makes its appearance as a phase
in the GW waveform at the fifth post-Newtonian order during the inspiral stage of a binary merger.
Various detection events could then measure or
set bounds on the TLN and thus the parameters (eg. equation of state parameters, theory couplings, etc.)
that it depends on (see for example \cite{LIGO1,LIGO2} for the most recent neutron binary (GW170817) data analysis performed by LIGO).

For black holes in general relativity, the TLN is conspicuously zero \cite{Porto,Damour}. 
In linearized perturbation theory, one can derive the differential equations which 
metric perturbations have to obey, and we can read off the TLNs from their 
large distance behavior. For the Schwarzschild black hole, the metric perturbations
can be analytically solved and upon imposing regularity at the horizon, one could
demonstrate the vanishing of its TLN. For neutron stars, we match the exterior spacetime 
to a suitable interior, and the TLN thus depends on the equation of state parameters
associated with the stellar interior. 

It is a natural and well-motivated question to explore how deviations from ordinary GR may
lead to the non-vanishing of TLNs, especially in view of the large amount of work dedicated to 
various theoretical models for its UV completion. 
Let us briefly point out recent related results on this issue. 
In \cite{Senatore}, black hole solutions
in an effective field theory framework encompassing various higher-order curvature extensions of the 
Einstein-Hilbert action were studied and it was found that they have non-vanishing TLNs. 
A similar work in \cite{Cai} found the same phenomenon for black holes in a theory with $R^3$ terms. 
In \cite{Cardoso}, black hole solutions in
Brans-Dicke theory, Chern-Simons gravity and Einstein-Maxwell theory were studied for their TLNs,
and non-vanishing quadrupolar and octopolar magnetic TLNs were found in the case of Chern-Simons gravity. 
The authors of \cite{Cardoso} also computed the TLNs of compact exotic objects which are not black holes such as
wormholes, boson stars, etc. which turn out to possess non-vanishing TLNs as well. Coupled with the increasing precision 
of GW detectors in the near future, this collection of recent 
results demonstrates that TLNs could be interesting and important indicators of new physics. 

In this paper, we study the TLNs for various black hole and wormhole solutions in a simple
context of warped compactification --- Randall-Sundrum theory \cite{Randall} in which our four-dimensional spacetime geometry
is that of a brane embedded in a five-dimensional spacetime which has $AdS_5$ as its vacuum. 
The conception of this model was originally motivated by the gauge 
hierarchy problem, and it has been actively studied at various levels, including its embedding in string theory, 
relevance for AdS/CFT, etc. In our opinion, this braneworld scenario is a simple and well-motivated backdrop where one could start to probe 
how extra
dimensions affect TLNs.  
In this case, the extra dimension is non-compact and gravity is localized on the brane by the warped geometry of the embedding spacetime. 

Most of this paper will be devoted to the electric, quadrupolar TLN commonly termed as $k_2$, since 
as shown in \cite{Flanagan}, it is the phenomenologically relevant one that appears at the fifth post-Newtonian order in the waveform. 
We will also present some essential and useful results for the general TLN associated with polar perturbations, but we leave axial-type 
or magnetic TLNs \cite{Pani:mag} for future work. The background braneworld solutions include black holes, wormholes and naked singularities which
were presented in the literature some time ago. They are exact solutions to a four-dimensional formulation of Randall-Sundrum model with an 
effective energy-momentum tensor (capturing the effects of the extra dimension) as first derived by Shiromizu, Maeda and Sasaki in \cite{Shiromizu}. The perturbation equations yield differential equations which appear difficult for analytic solutions but 
the computation of TLNs in this case can still proceed with appropriate series expansions about infinity and 
the metric singularities, provided the singular points of the differential equations are regular. 

The families of solutions we consider are all
parametrically connected to some well-defined limits of a particular black hole solution which share the same metric form with the Reissner-Nordstr$\ddot{\text{o}}$m solution but with a negative tidal charge. This `Tidal black hole' will be used as an anchor solution for us to check the consistency of various series solutions. In the vanishing charge limit, it reduces to the Schwarzschild metric \cite{Chamblin}. We should point out in \cite{Bose}, there was an attempt to compute the TLN for the Tidal black hole, but upon review (see details in Section \ref{Decoupled}),
we find that there is unfortunately a major error in that calculation. Nonetheless, we refer the reader to \cite{Bose} for a study of TLNs of braneworld stars \cite{Germani} which this paper does not cover. 

Here is a summary of our main results:
(i)TLNs of braneworld black holes and wormholes studied here are generically non-vanishing, 
(ii)they can be expressed for most cases as polynomials in the parameter that characterizes each family of solution,
(iii)indicial roots associated with near-horizon and asymptotic expansions display some level of universality across various families of solutions.

The paper is organized as follows. In Sections ~\ref{OnTLN} and ~\ref{OnRSL}, 
we furnish a review of the notions of TLNs and the effective field equations of the Randall-Sundrum braneworld model
respectively, establishing some conventions along the way. In Section ~\ref{OnPert}, we derive the perturbation equations and show
that we end up with either a homogeneous third-order ODE or an inhomogeneous second-order ODE to solve. In Sections 
~\ref{OnTBH} and ~\ref{Other}, we study each braneworld solution in detail, staying mostly focussed on the case of the quadrupolar TLN. In each case, we will briefly state relevant aspects of the causal structure of each solution following the relevant references which the reader can refer to for the Carter-Penrose (CP) diagrams and other elaborations. Finally, we end with some concluding remarks in Section ~\ref{Concl}. A couple of Appendices gather some explicit details on metric perturbation and indicial roots.

\section{On Tidal Love Numbers}
\label{OnTLN}

We begin by reviewing the notion of tidal love numbers (TLN) following
\cite{Poisson}. They are quantities which measure the effect of gravitational
tidal deformation due to some external companion or field on the object  
and its appearance in gravitational waveforms was
first studied seriously in \cite{Flanagan}. In Newtonian gravity, the TLN is a constant of proportionality between the tidal field applied 
to the body and the resulting multipole moment of its mass distribution. One can characterize the tidal field by the tidal moment 
\be
\label{tidalmoment}
\xi_{ab} (t) = - \partial_a \partial_b U_{ext},
\ee
where $U_{ext}$ is the Newtonian potential of some external body. 
This is evaluated at the body's CM and we are working in the object's local asymptotic rest frame
with the $x^a$ in \eqref{tidalmoment} being asymptotically CM centered Cartesian coordinates. 
It is a symmetric and tracefree tensor which can be covariantly described by the Weyl tensor\footnote{In Fermi
normal coordinates centered at $r=0$, $\xi_{ab} = R_{0a0b}$. See for example \cite{Misner}.  }. 
On the other hand, we define the quadrupole moment as 
\be
Q^{ab} = \int \rho(x) \, ( x^a x^b - \frac{1}{3}\delta^{ab} r^2 )\, d^3x,
\ee
where $\rho (x)$ is the mass density within the body. In the absence of the tidal field, we assume the body to be spherical 
and $Q^{ab}$ vanishes. In the presence of a weak tidal field,
from dimensional analysis, 
\be
Q_{ab} = -\frac{2}{3} k_2 R^5 \xi_{ab},
\ee
where $R$ is the body's radius and the rest are conventions. 
$k_2$ is the dimensionless tidal love number for a quadrupolar deformation and is the object of focus in this paper. 

More generally,
we could have tidal moments of higher multipole orders and higher powers of $x^a$. The set of tidal love numbers $k_l$ then measures the body's response. A useful way of characterizing this definition can be obtained  
from the expression of the Newtonian potential 
\be
U = \frac{M}{r} - \frac{1}{(l-1)l} \left[  1 + 2 k_l (R/r)^{2l+1} \right] \xi_L (t) x^L,
\ee
where $L \sim a_1 a_2... a_l$ is a multi-index that contains $l$ individual indices. The tidal moment
is now defined as 
$
\xi_L (t) = -\frac{1}{(l-2)!}  \partial_L U_{ext},
$ 
and the $l-$ pole moment of the mass distribution is the tensor
\be
Q^L = \int \rho x^{\langle L \rangle } d^3x, \qquad Q_L = - \frac{2(l-1)!}{(2l-1)!!} k_l R^{2l+1} \xi_L,
\ee
where $\langle L \rangle$ denotes the removal of its trace.
It is useful to formulate a working definition of the tidal love numbers from the following metric ansatz.
\bea
\label{metrictt}
g_{tt} &=& -1 + \frac{2M}{r} + \sum_{l \geq 2} 
\frac{2}{r^{l+1}}( M_l Y^{l0} + \ldots ) - \frac{2}{l(l-1)}r^l ( \xi_l Y^{l0} + \ldots ),\\
g_{t\varphi} &=& \frac{2J}{r} \sin^2 \theta + \sum_{l \geq 2} \frac{2}{r^l} ( \frac{S_l}{l} S^{l0}_\varphi + \ldots )
+ \frac{2r^{l+1}}{3l (l-1)} ( B_l S^{l0}_\varphi + \ldots ),
\eea
where $S^{l0}_\varphi = \sin  (\theta ) \partial_\theta Y^{l0}$. 
The `electric' and `magnetic' tidal love numbers are defined as 
\be
\label{k2E}
k^{(E)}_l =
-\frac{l(l-1)}{2M^{2l+1}}\frac{M_l}{\xi_{l}} ,\,\,\,
k^{(B)}_l =
-\frac{3l(l-1)}{2(l+1)M^{2l+1}}\frac{S_l}{B_{l}},
\ee
where we have absorbed a factor of $\sqrt{\frac{4\pi}{2l+1}}$ in $M_l, S_l$, with the two types of TLNs describing polar and axial perturbations.
Typically, we use gravitational perturbation theory to construct the metric describing the tidal deformations. 
One can parametrize static perturbations of spherically symmetric and static geometries is as follows (see for example \cite{Emparan}). Let $g_{\mu \nu} = g^{(bg)}_{\mu \nu} + h_{\mu \nu}$, then in the chart of $\{t, r, \theta, \phi \}$, we 
can write
\bea
h_{tt} &=& g^{(bg)}_{tt} \sum_l H^{(l)}_0 P_l, \,\,\,
h_{rr} = g^{(bg)}_{rr} \sum_l H^{(l)}_2 P_l, \cr
h_{\theta \theta } &=& r^2 \sum_l K^{(l)} P_l, \,\,
h_{\phi \phi } = r^2 \sin^2 \theta \sum_l K^{(l)} P_l, \,\,
h_{t\varphi } = \sum_l  h^{(l)}_0 P'_l \sin^2 \theta,
\eea
where $P'_l = \frac{dP_l (\cos \theta )}{d(\cos \theta )}$, and $P_l = P_l (\cos \theta )$ are Legendre polynomials used as basis functions for the angular dependence and due to the spherical symmetry of the ansatz, we suppress the degenerate azimuthal numbers. 
The above ansatz for static linearized perturbations was first studied by Thorne, Hartle and Campolattaro  \cite{Thorne,Thorne2}
and has since been frequently invoked in studies of TLN as well as quasi-Schwarzschild solutions \cite{Emparan} . 
For solutions
relevant for TLN, the metric is not asymptotically flat and is valid within a restricted domain that lies in the compact object's exterior and which captures the tidal effects of an external matter source.

In the remaining of our paper, we will focus mainly on $k^{(E)}_2 $- 
the dimensionless quadrupolar TLN which carries crucial phenomenological value due to its appearance as a tidal phase correction in the
gravitational waveform associated with the inspiral stage. Taking $l=2$ in
 \eqref{k2E}, we have
\be
k^{(E)}_2 = -\frac{1}{M^5} \frac{M_2}{\xi_{2}}, \qquad
k^{(B)}_2 = -\frac{1}{M^5} \frac{S_2}{B_2}.
\ee
We have suppressed other angular harmonics in our ansatz for the perturbation. Let us quickly note that
upon restoring them, we can write the metric component $g_{tt}$ in the form 
\be
\label{HindererAnsatz}
g_{tt} = -1 + \frac{2M}{r} + 
\frac{3 Q_{ij}}{r^3} \left( n^i n^j \right) - \xi_{ij} x^i x^j + \ldots
\ee
where $n^i = x^i/r$ is the unit vector and the quarupole moment $Q_{ij}$ is traceless. The $r^2$ term 
represents the tidal force and the $1/r^3$ term characterizes the compact object's response to the tidal force. 
To linear order in $\xi_{ij}$, the induced moment takes the form (see for example \cite{Hinderer} ) 
\be
Q_{ij} = - \lambda \xi_{ij}
\ee
for some constant $\lambda$. The dimensionless TLN $k^{(E)}_2$ is related by $k^{(E)}_2= \frac{3\lambda}{2M^5}$. 
Using the ansatz for $l=2$ and expanding in spherical harmonics, we can write
\be
Q_{ij} n^i n^j = \sum_{m=-2}^2 Q_m Y_{2m} , \,\,\,
\xi_{ij} n^i n^j = \sum_{m=-2}^2 \tilde{\xi}_m Y_{2m}.
\ee
Comparing between 
\eqref{HindererAnsatz} and \eqref{metrictt} yields 
\be
\lambda = - \frac{2M_2}{3\xi_2}, \qquad
k^{(E)}_2 = \frac{3\lambda}{2M^5}=-\frac{1}{M^5} \frac{M_2}{\xi_{2}}.
\ee
To extract $k^{(E)}_2$, we need the asymptotic series expansion of $h_{tt}$ from which we read
off  any non-vanishing pair of $r^2$ and $1/r^3$ terms.
Since we are seeking the induced response,
the relevant term of the form $1/r^3$ should vanish together with the tidal force term.
In the following, for each braneworld solution, we will focus on computing 
\be
\label{TLNbasic}
\lambda = \frac{1}{3} \frac{C_3}{C_2},
\ee 
where $C_3, C_2$ are the
coefficients of the $1/r^3, r^2$ terms in an asymptotic series expansion of $h_{tt}$. 
We will refer to $\lambda$ as the Tidal Love Number (TLN) from now on.

\section{On Randall-Sundrum theory with warped compactification }
\label{OnRSL}

In this section, we furnish some essential points concerning the background gravitational theory --- Randall-Sundrum theory with a 
brane of positive tension on which gravity is localized via curvature rather than compactification \cite{Randall}. 
We work with a five-dimensional bulk with $AdS_5$ as the vacuum and an energy-momentum tensor that is confined on the $3+1$D brane,
in which the effective cosmological constant can be set to vanish with a choice of brane tension. Following \cite{Shiromizu},
we find it useful to formulate the gravitational dynamics purely in terms of an effective four-dimensional theory which we review below. 
We will also derive how various metric components (including the perturbations) which are functions of the 3+1D brane-worldvolume coordinates appear in the line element obtained by expanding the metric about the brane.

We begin with a parent 5D metric ansatz that reads 
\be
\label{5Dmetric}
ds^2 = dy^2 + g_{\mu \nu} (x,y) dx^\mu dx^\nu,
\ee
where $y$ is a Gaussian normal coordinate that is orthogonal to the brane. 
In the neighborhood of $y=0$ where the confining brane lies, we can express the metric perturbation as
\be
g_{\mu \nu} (x,y) = g^{bg}_{\mu \nu} (x,y) + h_{\mu \nu} (x,y) = g^{bg}_{\mu \nu} (x,y) + h_{\mu \nu}(x,0) + 
h'_{\mu \nu} (x,0) y + \frac{1}{2} h''_{\mu \nu} (x,0) y^2 + \ldots,
\ee
where we denote $\partial_y$ with a prime for notational simplicity in this section. 
From \eqref{5Dmetric}, the extrinsic curvature describing the embedding of constant $y$ surfaces
reads
\be
\label{extrinsicA}
K_{\mu \nu} = \left(  \delta^\alpha_\mu - n^\alpha n_\mu \right)\left(  \delta^\beta_\nu - n^\beta n_\nu \right)\nabla_\alpha n_\beta = \frac{1}{2}  g'_{\mu \nu}.
\ee
We can fine-tune the brane cosmological constant to vanish, with
\be
\Lambda_5 = - \frac{\kappa^2_5 \lambda^2}{6}, \kappa^2_4 = \frac{1}{6} \kappa^4_5 \lambda,
\ee
where $\lambda$ is the brane tension, and $\kappa^2_{4,5} = 8\pi G_{4,5}$. 
In this paper, we will adopt this condition for simplicity. 
The ordinary 4D limit involves
taking $\kappa_5 \rightarrow 0, \lambda \sim \kappa^{-4}_5 \rightarrow \infty$ such that $\kappa_4$ is finite.  
Henceforth, we denote $\kappa_4$ simply as $\kappa$. 
The 5D field equations read
\be
G_{AB} = -\Lambda_5 g_{AB} + \kappa^2_5 \left( T_{AB} + T^{br}_{AB} \delta (y) \right).
\ee
For the energy content, we set 
\be
T_{AB} = 0, \,\,\,
T^{br}_{AB} = - \lambda g_{AB} \delta^A_\mu \delta^B_\nu.
\ee
Together with the fine-tuned vanishing of $\Lambda_4$, it can 
be shown using Gauss-Codazzi equations (see for example \cite{MaartensReview}) that
the field equations on the brane read
\be
\label{4Deffective}
G_{\mu \nu} = - E_{\mu \nu}, \,\,\,
E_{\mu \nu} = {C^\alpha}_{\beta \rho \sigma} n_\alpha n^\rho q^\beta_\mu q^\sigma_\nu = 
K_{\mu \alpha} {K^\alpha}_\nu - \partial_y K_{\mu \nu} - \frac{\Lambda_5}{6}g_{\mu \nu}.
\ee
Apart from the continuity of metric across $y=0$,
the Israel junction conditions impose the discontinuity of the extrinsic curvature to be
\be
K^+_{\mu \nu} - K^-_{\mu \nu} = - \kappa^2_5  \left(  T^{br}_{\mu \nu} - \frac{1}{3}T^{br} g_{\mu \nu} \right)
=-\frac{1}{3}\kappa^2_5 \lambda g_{\mu \nu},
\ee
which implies that on the brane,
\be
K_{\mu \nu} = - \frac{1}{6} \kappa^2_5 \lambda g_{\mu \nu} = \frac{1}{4} K g_{\mu \nu}.
\ee
By virtue of the metric ansatz, we also have \eqref{extrinsicA} which leads to the following useful relation on the brane
\be
\label{metricB}
g'_{\mu \nu} (x,0) = \frac{1}{2}K g_{\mu \nu} (x,0).
\ee
With the metric perturbation switched on, the scalar $K = -\frac{2}{3} \kappa^2_5 \lambda$ remains invariant and 
specifies the constraint on the brane
\be
g^{\mu \nu} (x,0) g''_{\mu \nu} (x,0) = K^2. 
\ee
The perturbations have to satisfy (from \eqref{metricB})
\be
h'_{\mu \nu} (x,0) = \frac{1}{2} K h_{\mu \nu} (x,0).
\ee
Under the perturbation, we find 
\be
\delta E_{\mu \nu} = -\frac{1}{2}h''_{\mu \nu} - \frac{1}{6}\Lambda_5 h_{\mu \nu} + \frac{1}{2} h'_{\mu \alpha} 
{K^\alpha}_\nu + \frac{1}{2}{K_\mu}^\beta h'_{\beta \nu}.
\ee
When restricted on the brane, we find 
\be
\label{deltaE}
\delta E_{\mu \nu} (x,0) = -\frac{1}{2} h''_{\mu \nu} (x,0) + \frac{3}{16} K^2 h_{\mu \nu} (x,0),
\ee
with a slightly different background relation
\be
\label{E}
E_{\mu \nu} (x,0) = -\frac{1}{2} g''_{\mu \nu} (x,0) + \frac{1}{8} K^2 g_{\mu \nu} (x,0).
\ee
We note that equations \eqref{deltaE} and \eqref{E} imply that up to second-order in $y$, we have
the following expansion of the metric components. 
\bea
\label{p11}
g^{bg}_{\mu \nu} &=& g^{bg}_{\mu \nu} (x,0) + \frac{1}{2}K g^{bg}_{\mu \nu} (x,0) y + \left[ 
\frac{1}{8}K^2 g^{bg}_{\mu \nu} (x,0) - E_{\mu \nu} (x,0) 
\right]y^2 + \ldots ,\\
\label{p22}
h_{\mu \nu} &=& h_{\mu \nu} (x,0) + \frac{1}{2}K h_{\mu \nu} (x,0) y + \left[ 
\frac{3}{16}K^2 h_{\mu \nu} (x,0) - \delta E_{\mu \nu} (x,0) 
\right]y^2 + \ldots 
\eea
In this paper, we will work with various classical solutions in an effective 3+1D description as backgrounds. 
We will not seek their 4+1D completion beyond what they imply for the bulk extension via \eqref{p11} and \eqref{p22}. Although it has proven difficult to find the full
solution in the bulk, the effective 4D field equations were shown in \cite{Shiromizu} to be fully consistent.\footnote{In \cite{Shiromizu}, it was shown that one can integrate into the bulk by complementing the effective 3+1D description by two more differential equations involving the Lie derivative of $E_{\mu \nu}$ and the Weyl tensor as explained in the  Appendix of \cite{Shiromizu}.
We also refer the interested reader to \cite{Seahra} which contains discussions of how Campbell-Magaard type embedding theorems in differential geometry imply the existence of bulk solutions from solutions in Shiromizu-Maeda-Sasaki formulation. }

\section{On the perturbation equations}
\label{OnPert}

From \eqref{4Deffective}, we see that $E_{\mu \nu}$ should be interpreted as an energy-momentum tensor satisfying the Bianchi identity. We proceed by adopting the following ansatz for it 
\be
-\frac{1}{\kappa^2_4}E_{\mu \nu} = \rho \left( U_\mu U_\nu + \frac{1}{3}h_{\mu \nu} \right) + q_{(\mu} U_{\nu )} 
+\Pi \left( \frac{1}{3}h_{\mu \nu} - r_\mu r_\nu \right)
\ee
where $r_\mu, U_\mu$ are unit radial and time-like vectors respectively, $h_{\mu \nu} = g_{\mu \nu} 
+ U_\mu U_\nu$ and the quantities $\rho, \Pi$ are the effective density and stress induced on the brane. We also
set $\kappa_4=1$ from now on. 
This ansatz has turned out to be very useful in seeking classical solutions (see for example \cite{MaartensReview}
for an extensive review). 
For the solutions we consider, we take $q_\mu$ to vanish and $\rho = \rho(r), \Pi = \Pi (r)$ 
which generally correspond to spherically symmetric and static geometries.  We adopt the metric ansatz
\be
ds^2 = -f(r) dt^2 + g(r) dr^2 + r^2 \left( d\theta^2 + \sin^2 \theta d\phi^2 \right).
\ee
The Bianchi identity can be further simplified to be a single ODE in the radial coordinate
\be
\label{Bianchi}
\rho' + \frac{f'(r)}{f(r)} \left( -\Pi + 2\rho \right) -2\Pi' - \frac{6}{r} \Pi = 0.
\ee
The background field equations read
\be
\label{Backgd}
G^t_t = = - \rho, \,\,\,
G^r_r = - \frac{1}{3} (2\Pi - \rho ),\,\,\,
G^\theta_\theta = G^\phi_\phi =  \frac{1}{3} (\Pi + \rho ).
\ee
In terms of the functions $f,g$,
\be
G^t_t = \frac{
g(r) - g(r)^2 - r g'(r)} 
{r^2 g(r)^2}, \,\,
G^r_r = 
\frac{f(r) (1-g(r)) + r f'(r) }{r^2 f(r) g(r)},
\ee
\be
\label{Backgd2}
G^\theta_\theta =G^\phi_\phi = \frac{
-r g(r) (f'(r))^2 - 2 f^2(r) g'(r) + f(r) \left(  -r f'(r) g'(r) + 2g(r) (  f'(r) + rf''(r)) \right)}
{4r f^2(r) g^2 (r)
}.
\ee
We express the perturbations in the basis of Legendre polynomials $P_l (\cos \theta )$. 
\bea
&&f(r) \rightarrow f(r) (1 + \sum_l H^{(l)}_0 (r) P_{l} ), \,\,\, g(r) \rightarrow g(r) (1 + \sum_l H^{(l)}_2 (r) P_{l}   ), \cr
&&g_{\theta \theta} = g_{\phi \phi}/\sin^2 \theta = r^2 \rightarrow r^2 \sum_l K^{l} (r) P_{l},\cr
&& \rho (r) \rightarrow \rho (r) +\sum_l  \delta \rho^{l} (r) P_{l}, \,\,\, \Pi (r) \rightarrow \Pi (r) + \sum_l \delta \Pi^{l} (r) P_{l}.
\eea
Also the Bianchi identity translates into an ODE for the perturbations of $\rho, \Pi$ as follows. From \eqref{Bianchi}
\be
\label{B1}
\delta \rho' + 2 \frac{f'(r)}{f(r)}  \delta \rho + (2\rho - \Pi )H'_0 = 2 \delta \Pi' + \delta \Pi \left( \frac{f'(r)}{f(r)} + \frac{6}{r} \right).
\ee
Since $\rho, \Pi$ parametrize the 5D graviton perturbations, we take $\delta \rho, \delta \Pi$ to be of the 
same fluctuation order as the 4D metric perturbations $\{H_0(r), H_2(r), K(r) \}$.

%------------------------------------------------------------------------

\subsection{Decoupling of angular modes and a third-oder ODE}
\label{Decoupling}

In the following, we show that the angular polar modes in the basis of Legendre polynomials $P_{l} (\cos \theta )$ decouple in the linearized equations, and derive
the ODE that determines $H_0(r), H_2 (r), K(r)$. From now on, we omit the superscript $(l)$ on these functions and their dependence on $r$ for notational simplicity. 

From $\delta G^r_\theta=0$, we find the relation
\be
K' = - H_0'  + \frac{1}{r} (H_0 + H_2 ) + \frac{f'}{2f} (H_2 - H_0 ),
\ee
whereas from $\delta G^\theta_\theta - \delta G^\phi_\phi$, we obtain
\be
\delta G^\theta_\theta - \delta G^\phi_\phi = \frac{1+l}{r^2 \sin^2 \theta} (H_0  + H_2  ) 
\left[ 
(4+l+(2+l)\cos (2\theta ) ) P_l - 2 (5+2l) \cos (\theta ) P_{1+l} + 2 (2+l)P_{2+l} 
\right],
\ee
which implies for $l \geq 2$ that\footnote{For $l=0,1$, the RHS vanishes identically, but $K$ decouples from all equations and can be solved by the same quadrature once we solve two coupled ODE in $H_0, H_2$.} 
\be
\label{Pertb}
H_2  = -H_0, \qquad K' = -H'_0 - \frac{f'}{f}H_0.
\ee
We find that $K$ in eliminated in $\delta G^r_r - \delta G^t_t$ which reads
\bea
\label{diffG}
\delta G^r_r - \delta G^t_t &=&  - \frac{1+l}{r^2 \sin^2 \theta}H_0  \left[
(3+2l) \cos (\theta ) P_{1+l} - (2+l) P_{2+l}
\right]  \cr
&& \qquad + \frac{P_l}{2r^2 f^2 g^2 } \Bigg[  -3r^2 g H_0  f'^2 + r f \bigg(
-rH_0 f' g' + g (6H_0 f'  \cr
&&\qquad + rf' H'_0  + 2r H_0  f'' )
\bigg)  + f^2  \bigg[   
(1+l)(2+l+l \cos (2\theta ) )\frac{g^2 H_0 }{\sin^2 \theta} \cr
&& \qquad +r g' (2H_0  -r H'_0 ) + 2r g (2H'_0 + rH''_0  )
   \bigg]
\Bigg].
\eea
The appearance of $P_{1+l}, P_{2+l}$ may suggest possible couplings among different polar modes, but at this point we invoke
Bonnet's recursion formula 
$$
(l+2)P_{2+l} - (2l+3)\cos (\theta ) P_{l+1} + (l+1)P_l = 0
$$
to simplify eqn. \eqref{diffG} to be
\bea
\label{diffG1}
\delta G^r_r - \delta G^t_t &=&  - \frac{l(l+1)}{r^2 }P_l H_0  + \frac{P_l}{2r^2 f^2  g^2} \Bigg[  -3r^2 g H_0  f'^2 + r f \bigg(
-rH_0  f' g' \cr
&&\,\, + g (6H_0  f'  + rf' H'_0  + 2r H_0  f'' )
\bigg)  + f^2  \bigg[ r g' (2H_0 
-r H'_0 ) + 2r g (2H'_0  + rH''_0  )
   \bigg]
\Bigg]. \nonumber \\
\eea
Thus we see there is decoupling of the polar modes since the angular dependence lies purely in $P_l (\cos (\theta ))$.
This should be equated to $\frac{2}{3} ( 2\delta \rho - \delta \Pi ) P_l $. We are left with the tracelessness condition which however
involves $K(r)$. 

We now use the field equations and Bianchi identity to write down a single ODE for $H_0$. 
First, from  \eqref{Backgd2} and \eqref{Pertb}, the field equations imply in particular that
\be
\label{Relation}
\delta G^\theta_\theta = \delta G^\phi_\phi = \frac{1}{2 r f g^2} H_0  (g f)' P_l = \frac{1}{3} ( \delta \Pi + \delta \rho ) P_l,
\ee
which allows one to express $\delta \Pi$ in terms of $\delta \rho$. This equation is identical for all $l\geq2$. Substituting \eqref{Relation} into 
\be
\label{B2}
\delta \rho' + 2 \frac{f'}{f}  \delta \rho + (2\rho - \Pi )H'_0 = 2 \delta \Pi' + \delta \Pi \left( \frac{f'}{f} + \frac{6}{r} \right),
\ee
we find the following third-order ODE
\be
\label{ThirdOrderODE}
C_3 H'''_0 + C_2 H''_0 + C_1 H'_0 + C_0 H_0 = 0,
\ee
where 
\bea
C_3 &=& \frac{r^2 f}{2g}, \,\,\, 
C_2 = \frac{r}{4g^2} \left[ 
3rgf' + f (8g-3rg')
\right], \cr
C_1 &=& \frac{1}{4fg^3} \bigg( -3r^2 g^2 f'^2 + rfg (-3rf'g' + g(16f'+3rf'')) \cr
&&\qquad -f^2 \left[ -4g^2 + 2l (l+1)g^3 - 2r^2 g'^2 + r^2 g g'' \right] \bigg), \cr
C_0 &=& \frac{1}{4f^2 g^3} \bigg( 3r^2 g^2 f'^3 + rgff' ( 3rf'g'-2g(4f'+3rf'') ) \cr
&& \qquad + 4f^3 (-2r g'^2 + g(2g'+rg'') ) - 
f^2 \bigg[ 2l (1+l) g^3 f' - 2r^2 f'g'^2  \cr
&& \qquad + rg (3rg'f''+f'(8g'+rg'') ) - 2g^2 (6f'+r(6f''+rf''')) \bigg] \bigg).
\eea

\subsection{The decoupled case $\Pi = 2\rho, f g=1$ }
\label{Decoupled}

There is a distinguished case of $$\Pi = 2\rho = -\frac{q}{r^4}, f =1/g = 1-\frac{2m}{r} +\frac{q}{r^2},$$ 
which corresponds to the important example of the Tidal black hole. We will study this solution in detail in Section~\ref{OnTBH} .
In this case we find that $\delta \rho$ decouples
from the differential equations. From \eqref{Relation}, we obtain 
\be
\label{Tidalrho}
\delta \rho  = -\delta \Pi,
\ee
which, upon solving \eqref{B2}, leads to $\delta \rho \sim \frac{1}{r^2 f}$.

At this point, we should point out that in \cite{Bose}, while studying this black hole solution, 
the authors unfortunately took the (correct) relation $\Pi = 2\rho$ to imply the (incorrect) relation 
$\delta \Pi = 2\delta \rho$. Although it is true that there is no a priori constraints between $\rho, \Pi$, the 
linearized perturbations of these functions have to satisfy the field equations and Bianchi identity for consistency. 
and \eqref{Relation} simply leads to \eqref{Tidalrho} for this solution (without imposing any further constraints by hand, for example by assuming $\delta \rho = C \delta \Pi$ for some constant $C$). 

In this case, equation \eqref{diffG1} simplifies to read 
\be
\label{DecoupledODE}
H''_0 + \frac{P(r)}{r} H'_0 + \frac{Q(r)}{r^2} H_0 =-l(l+1) \alpha \frac{r^2}{(r (r-2 m)+q)^2},
\ee
where $\alpha$ parametrizes the arbitrary constant for $\delta \rho$,\footnote{The factor of $l(l+1)$ is introduced for 
a slightly simpler notations for related computations in Section~\ref{OnTBH}.}
and 
\bea
P(r) &=& -\frac{2 r (m-r)}{r (r-2 m)+q}, \cr
Q(r) &=& -\frac{q r \left(\left(l^2 + l - 2\right) r-4 m\right)+r^2 \left(-2 l (l+1) m r + l (l+1) r^2+4 m^2\right)+2 q^2}{ (r (r-2 m)+q)^2}.
\eea
In Section~\ref{OnTBH}, we will discuss the near-horizon and asymptotic series solutions to \eqref{DecoupledODE}, focussing mostly
on the more phenomenologically relevant $l=2$ case.

%-------------------------------------------------------------------------------------------------
\section{On the Tidal Black Hole}
\label{OnTBH}

In this section, we study the tidal deformation of the following black hole solution first presented in \cite{TidalBH} with metric 
\be
f(r) = g(r)^{-1} = 1 - \frac{2m}{r} + \frac{q}{r^2}, \qquad q<0.
\ee
This is of the usual Reissner-Nordstr$\ddot{\text{o}}$m form but we note that the corresponding energy-momentum tensor is of course not of Maxwell's theory. The parameter $q$ is known as the tidal charge parameter and is of the negative sign which 
can be interpreted as a strengthening of the gravitational field. 
We will henceforth refer to this solution as the `Tidal Black Hole'. 
There is a regular horizon at 
\be
r=r_h = m \left[ 1  + \sqrt{1 - \frac{q}{m^2}}
\right].
\ee
The other zero of $f$ lies at a negative $r_- =  m \left[ 1  -  \sqrt{1 - \frac{q}{m^2}} \right]$. 
The background fields read
\be
G^t_t = G^r_r= -G^\phi_\phi=-G^\theta_\theta=
-\frac{q}{r^4},\qquad
\rho = -\frac{q}{r^4} = \frac{ \Pi}{2}.
\ee
Setting $l=2$ in \eqref{DecoupledODE}, we obtain the equation 
\be
\label{TidalODE}
H''_0 + \frac{P (r)}{r} H'_0 + \frac{Q(r)}{r^2} H_0 = \frac{-6\alpha r^2}{(q+r^2-2mr)^2} \equiv \mathcal{S} (r),
\ee
where
\be
P(r) = - \frac{-2r (m-r)}{q+r(r-2m)}, \,\,\,
Q (r) = \frac{
(-2 (q^2 + 2 q r (-m + r) + r^2 (2 m^2 - 6 m r + 3 r^2)))
}
{(q + r (-2 m + r))^2}.
\ee
By the general theory of ODEs, \eqref{TidalODE} has two independent homogeneous solutions and 
a particular solution. Their analytic forms are not known to us, 
yet in the $q=0$ limit, exact solutions are known. 
In the following, we study their near-horizon
series expansion in parameter $x=(r-r_h)/m$ and asymptotic series expansion in parameter $u=m/r$. 
They can be used to compute the TLN for the Tidal black hole that turns out to be a simple function of mass and tidal charge under certain assumptions for their regularity at the horizon.

%-------------------------------------------------------------------
\subsection{Expansion about $r = r_h$}
\label{ExpHor}
We first carry out an expansion about $r=r_h$, and impose regularity at the horizon. 
Let $x=(r-r_h)/m$ and recast the ODE in variable $x$, 
writing \eqref{TidalODE} as
\be
H''_0 + \frac{\tilde{P} (x)}{x} H'_0 + \frac{\tilde{Q} (x)}{x^2} H_0 = m^2 \mathcal{S} (x)
\ee
where $\tilde{P} (x) = x P(x)/(x+r_h), \tilde{Q} (x) = x^2 Q(x)/(x+r_h)^2$. 
We find $x=0$ to be a regular point with 
$$
\tilde{P} (0) = 1, \, \tilde{Q} (0) = -1.
$$
The indicial equation now reads
\be
\mathcal{R}(\mathcal{R}-1)+\mathcal{R} \tilde{P}(0) + \tilde{Q}(0) = 
\mathcal{R}(\mathcal{R}-1) +\mathcal{R}-1 = 0, 
\ee
which has roots being $\pm 1$ independent of $q$. 
The two homogeneous solutions are thus of the form
\bea
\label{HorReg}
\tilde{S}_r (x,q) &=& x \left( 1 + a_1 x + a_2 x^2 + a_3 x^3 + \ldots \right), \\
\label{HorDiv}
\tilde{S}_d (x,q) &=& \mathcal{N} \tilde{S}_r (x,q) \log (x) + \frac{1}{x} \left( 1 + b_1 x + b_2 x^2 + b_3 x^3 + \ldots \right).
\eea
Since $\tilde{S}_d(x,q)$ diverges as $x\rightarrow 0$, this implies that $g_{rr}$ is divergent at the horizon - 
a feature which nullifies the series as part of the general solution regular at horizon. We also find the particular solution of the following form
\be
\label{particularTBH}
\tilde{S}_p (x,q) = C_0 + \left(C_1 + \mathcal{K}_1 \log (x) \right) x + \left(C_2 + \mathcal{K}_2 \log (x) \right) x^2 + \ldots
\ee
In Appendix \ref{AppA1}, we collect several explicit expressions for some of the series coefficients in \eqref{HorReg}, \eqref{HorDiv}
and \eqref{particularTBH} as useful references. The coefficients $a_i$ in $\tilde{S}_r (x,q)$ are 
uniquely determined as functions of $q,m$ while in $\tilde{S}_d (x,q)$, the coefficients $b_i, i \geq 2$ are 
defined up to an arbitrary multiple of $\tilde{S}_r(x,q)$ being added to $\tilde{S}_d (x,q)$. 
The particular solution is defined such that it vanishes in the $\alpha \rightarrow 0$ limit.

The series expressions $\tilde{S}_r (x,q), \tilde{S}_d(x,q)$ and $\tilde{S}_p (x,q)$ are the near-horizon series expansions of the two
 independent 
homogeneous solutions and the particular solution of \eqref{TidalODE} up to their 
linear combinations, 
with leading-order terms $\sim x^{1}, x^{-1}, x^{0}$ respectively. In the $q=0$ limit, these series solutions read
\bea
\label{HorizonTid2}
\tilde{S}_r (x,q=0) &=& x \left( 1 + \frac{1}{2}x  \right) = \left(\frac{r}{m}-2 \right)\frac{r}{2m}, \cr
\tilde{S}_d (x,q=0) &=& -3 \tilde{S}_r (x,q=0) \log \left( x \right) + \frac{1}{x} \left( 1 -\frac{5}{2} x + \frac{13}{4} x^3 + \ldots \right), \cr
\tilde{S}_p (x,q=0) &=&\alpha \left( 1 + (\log (x)) x + \left( - \frac{13}{12} + \frac{1}{2}\log (x) \right) x^2 - \frac{5}{48} x^3 + \ldots \right)
\eea
We can compare them with the Schwarzschild case in ordinary GR for which there is an exact solution for 
$H_0$ first found by Hinderer to be  
\be
\label{H0Hind}
H_0 = c_1 \frac{5}{8} \left(  \frac{r}{m}  \right)^2 \left(1-\frac{2m}{r} \right) \left[ 
-\frac{
m(m-r)(2m^2 + 6mr - 3r^2)
}{
r^2(2m-r)^2
}
-\frac{3}{2}\log(1-\frac{2m}{r})
\right] +c_2 \left( \frac{r}{m} \right)^2 (1-\frac{2m}{r}).
\ee
where $c_1, c_2$ are arbitrary constants.
Expanding about the horizon $r_h =2m$, 
we find that \eqref{H0Hind} is equivalently 
\be
\label{H1Hind}
H_0 (x) = c_1 \frac{5}{8}
 \left( \tilde{S}_d(x,q=0) - \left( \frac{13}{4} -3 \log (2) \right) \tilde{S}_r (x,q=0) \right) 
+ 2c_2 \tilde{S}_r (x,q=0) 
\ee
For convenience, we define 
\be
\tilde{y}_d (x,q) = \frac{5}{8} \tilde{S}_d (x,q) - \left( \frac{13}{4} -3 \log (2) \right) \tilde{S}_r (x,q), \,\,\,
\tilde{y}_r (x,q) = 2 \tilde{S}_r (x,q), \,\,\, \tilde{y}_p (x,q) = \tilde{S}_p (x,q),
\ee
after which \eqref{H1Hind} can be expressed as 
\be
H_0 (x) =  c_1  \tilde{y}_d(x,q=0) + c_2 \tilde{y}_r (x,q=0).
\ee
At this point, we note that the TLN is computed from the solution that is regular at the horizon. 
Discarding the solution $\tilde{y}_d (x,q=0)$ which is divergent at the horizon, we are left with $\tilde{y}_r (x,q=0)$ which can be used to show that the Schwarzschild black hole in ordinary GR has vanishing TLN. We note that the particular solution vanishes in the ordinary GR limit, since $\alpha$ parametrizes fluctuations of the matter density that effectively descend purely from the 5D brane embedding (and not from 4D matter). As a side note, for the higher values of $l$, we find that we have the same indicial equation as in the $l=2$ case, with the roots being $\pm 1$. There is also a particular solution of the same form as in the $l=2$ case. This once again implies that there are two regular branches of solutions, with one of them relevant for capturing the tidal response. Although the expansion about the horizon allows us to match the general form of regular solution to the Schwarzschild case in the limit of ordinary GR, we need the expansion about infinity to study the TLN since the latter is most conveniently read off from such a series. 
 
For each of the two independent functions of $r$ in \eqref{H0Hind} or equivalently $\tilde{y}_{\{d,r\} } (x,q=0)$, 
we can also perform a series expansion about infinity in the parameter $u=m/r$. 
We find
\bea
\label{Asym1}
\tilde{y}_d (x,q=0) &=& u^3 + 3u^4 +\frac{50}{7}u^5 + \ldots, \\
\label{Asym2}
\tilde{y}_r (x,q=0) &=& \frac{1}{u^2} - \frac{2}{u}.
\eea
In the next Section, we will solve for the asymptotic series expansion and compare
the solutions against \eqref{Asym1} and \eqref{Asym2} in the $q=0$ limit.

\subsection{Expansion about $r=\infty$}
\label{ExpInf}
We now perform a series expansion about $r=\infty$ in order to pick up any non-vanishing TLN. 
After switching to dimensionless parameter $u = m/r$, we rewrite the ODE as
\be
H''_0 + \frac{2-P(u)}{u} H'_0 + \frac{Q(u)}{u^2} H_0 = \frac{m^2 \mathcal{S}(u)}{u^4},
\ee
and find the expansion behavior
\bea
2-P(u) &=& -2u + (2q-4)u^2 + \ldots, \cr
Q(u) &=& -6-12 u + \ldots, \cr
\frac{m^2 \mathcal{S} (u)}{\alpha u^4} &=& \frac{1}{u^2} + \frac{4}{u} + (12 - 2q) + \ldots,
\eea
which reveal $r=\infty$ or $u=0$ to be a regular point. The indicial equation reads 
\be
\mathcal{R}(\mathcal{R}-1)+\mathcal{R} (2-P(0)) + Q(0) = \mathcal{R}(\mathcal{R}-1) -6 = 0,
\ee
which has roots 3 or $-2$. This gives the leading-order indices for the 
two independent homogeneous solutions which we find to be
\bea
\label{Yd}
y_d (u,q) &=& u^3 \left( 1 + 3 u + \frac{1}{7m^2} (50m^2 - 7q) u^2 + \frac{1}{42}\left(660-217\frac{q}{m^2}\right) u^3 + \ldots \right), \\
\label{Yr}
y_r (u,q) &=& 
\frac{1}{u^2} - \frac{2}{u} + \frac{2q}{3m^2} + \frac{2qu}{3m^2} - \frac{2q^2 u^2}{3m^4}
-\frac{16}{15m^4} (q^2 \log (u) ) u^3 + \frac{2q^2}{45m^4}\left(31  + 15 \frac{q}{m^2} - 72 \log (u) \right) u^4 + \ldots \nonumber \\
\eea
We also find the particular solution to be of the following form
\be
\label{Parti}
\frac{y_p (u,q)}{\alpha} = 1+2 u+(4-\frac{q}{m^2}) u^2+\left(\frac{q^2}{m^4}-8\right) u^4+\mathcal{O}\left(u^5\right).
\ee 
Let's first consider their $q=0$ limit :
\bea
\label{Asy1}
y_d (u,q=0) &=& u^3 (1 + 3u + \frac{50}{7}u^2 + \frac{110}{7}u^3 + \ldots ), \\
\label{Asy2}
y_r (u,q=0) &=& \frac{1}{u^2} (1-2u ), \\
\label{Asy3}
y_p(u,q=0) &=& \alpha \left( 1 + 2u + 4u^2  - 8 u^4 + \ldots  \right).
\eea
We find that \eqref{Asy1} and \eqref{Asy2} are precisely \eqref{Asym1} and \eqref{Asym2} respectively.
From the preceding section, we see that the near-horizon expansion 
of $y_d (u,q)$ can be expressed in the form 
\be
\label{ydproof}
y_d (u,q) = \kappa_{(1)} (q) \left( x + \ldots \right) + \kappa_{(0)} (q) \left(1 + \ldots \right) + \kappa_{(-1)} (q) 
\frac{1}{x}  \left( 1 + \ldots \right),
\ee
where we have kept only the leading order terms in $\tilde{y}_r (x,q), \tilde{y}_p (x,q) $ and $\tilde{y}_d (x,q)$,
and the $\kappa$'s are constants with possibly $q-$dependence. The $q=0$ limit implies that $\kappa_{(-1)} (0) = 1$ and since $\kappa_{(-1)} (q)$ is non-vanishing for a general $q$, $y_d (u,q)$ diverges at the horizon. 

For $y_r (u,q)$, we cannot however invoke the above argument to prove the absence of $\tilde{S}_d(x,q)$ or $\tilde{y}_d (x,q)$ in its near-horizon expansion, and thus prove its regularity. 
In the absence of analytic solutions, we proceed by assuming that $y_r (u,q)$ is regular at the horizon, noting the following:
\begin{enumerate}[(a)]
\item A direct way to prove whether $y_r (u,q)$ is regular at horizon is to study whether one could perform some resummation of 
the infinite series in \eqref{Yr}. For example, if we regard \eqref{Yr} as the asymptotic series of a function of $q$ and $r$ that is also perturbative in $q/m^2$ (which is our physical regime of interest), then one could write it as a series in $q$. Generically, we expect each coefficient to be an infinite series in $u$ and for \eqref{Yr}, this turns out to be the case for all $q^k, k >1$ terms but not for the linear term. 
At order $\mathcal{O} (q^2)$, $y_r (u,q)$ can be written as a finite sum of terms which reads
\be
\label{trunAsymp}
y_r (u,q) = \frac{1}{u^2} - \frac{2}{u} + 
\frac{2q}{3m^2} (1+u) + \mathcal{O}(q^2).
\ee
This is regular and in fact vanishes at the horizon $r_h =2m - \frac{q}{2m} + \mathcal{O}(q^2)$. Thus, 
at order $\mathcal{O}(q^2)$, we have 
\be
\label{rtruncated}
y_r (u,q) = (1+C_r q) \tilde{y}_r (x,q),
\ee
with the RHS understood to be evaluated at linear order, the $q=0$ limit fixing the unity in the prefactor and $C_r$ is an undetermined constant.\footnote{In principle we can compute $C_r$ if we expand $\tilde{y}_r (x,q)$ in $q$ to linear order, but we find this difficult with $x=(r-r_h)/m$ being an infinite series in $q$.} Although this is compatible with the conjecture that $y_r (u,q)$ is regular at horizon, unfortunately it doesn't rigorously
complete a proof for its regularity since it doesn't extend naturally to higher orders. 
Nevertheless, it indicates that functions of $r$ that are divergent at horizon could enter into the $q$-series of $y_r (u,q)$ starting only at or above order $\mathcal{O}(q^2)$, if at all. 

\item  Suppose that contrary to our assumption, $y_r (u,q)$ diverges at the horizon, then the regular solution we seek 
 is a suitable linear combination of $y_r(u,q)$ and $y_d (u,q)$. 
For the latter, the strict $q=0$ limit implies that its near-horizon expansion is of the form 
\be
y_d (u,q)=(1+ \mathcal{O}(q)) \tilde{y}_d (x,q) + \mathcal{O}(q) \tilde{y}_r (x,q) + \mathcal{O}(q) \tilde{y}_p (x,q).
\ee
Also, the preceding point in (a) implies that we can write
\be
y_r (u,q) = (1+C_r q + \mathcal{O}(q^2)) \tilde{y}_r (x,q) + \mathcal{O}(q^2) \tilde{y}_d (x,q) + \mathcal{O} (q^2) \tilde{y}_p (x,q).
\ee
It is then straightforward to see that at order $\mathcal{O}(q^2)$, the linear combination of $y_d (u,q)$ and $y_r (u,q)$ that reduces to $y_r (u,q=0)$ 
in the $q=0$ limit is of the form 
\be
\label{divR}
(1+N_r q) \left( y_r(u,q) + \mathcal{O}(q^2) y_d (u,q) \right),
\ee
where $N_r$ is an arbitrary constant, since the term $\tilde{y}_d (x,q)$ must be cancelled away. Thus, 
at linear order, \eqref{divR} is of the same form as 
the RHS of \eqref{rtruncated}.

\item  As we reviewed in Section 4, in the effective 4D 
gravitational theory, the Tidal black hole is a solution to the case where the 
energy density $\rho$ and stress $\Pi$ take the form $\Pi = 2\rho =-\frac{q}{r^4}$. 
If $\alpha=0$ (excluding the particular solution), then $\delta \rho = \delta \Pi = 0$ and the perturbations
are vacuum perturbations. Now if $y_r (u,q)$ diverges at the horizon, then formally, 
the near-horizon limit and $q\rightarrow 0$ limit do not commute. 
Physically, this translates to a picture where adiabatically turning on a small tidal charge
induces a large (vacuum) perturbation near the horizon which we find somewhat difficult to
understand, at least by naive intuition.\footnote{On a slightly related context, we note that the tidal black hole has been found to
be stable against different types of perturbations (see for example \cite{Tosh}). }
The regularity of $y_r (u,q)$ at the horizon may appear to be more physically 
reasonable than its converse.

\end{enumerate}

Henceforth, we will assume that $y_r (u,q)$ is regular at the horizon, our choice motivated by the points 
(a) and (c) discussed above and urged by feasibility of computation.
For future work, it will be important to
study our assumption more rigorously, by for example studying the $q$-series expansion of the 
ODE \eqref{TidalODE} and terms higher-order in $q$. If all of these can be appropriately resummed, we could then know,
beyond the linear order expressed in \eqref{trunAsymp}, whether it is genuinely regular at the horizon. 

Finally, we note that the particular solution does not contain the tidal moment ($r^2$-term) 
although there is a non-vanishing quadrupolar moment due to the 5D embedding.
Since we are considering the induced response of the body to some tidal force,
when reading off the TLN, we consider purely
$y_r (u,q)$ which contains the tidal moment term and possibly some induced quadrupolar moment. 

Thus, taking $\alpha=0$, we have the asymptotic series expansion
\be
\label{TidalRegF}
H^{(reg)}_0 = \mathcal{C} \left(  \left(  \frac{r}{m} \right)^2 + \frac{16}{15}q^2 \log \left(  \frac{r}{m}   \right) \frac{m^3}{r^3} - 2\frac{r}{m} + \frac{2q}{3m^2} + \frac{2q}{3mr}
-\frac{2q^2}{3m^2r^2} + \ldots + \sum_{k=4} \left( \tilde{C}_k + \tilde{D}_k \log (u) \right) u^k \right),
\ee
where $\tilde{C}_k, \tilde{D}_k$ are constant coefficients which can be computed straightforwardly whenever required. 
This expression does not contain $1/r^3$ term on its own, 
but taking into account $f(r) = 1- \frac{2m}{r}+\frac{q}{r^2}$, with $h_{tt} = f(r) H^{(reg)}_0$, we find from 
\eqref{TidalRegF} and \eqref{TLNbasic} that 
\be
\label{lambdaTidal}
\lambda = \frac{2mq^2}{3}.
\ee
This is our first example of a braneworld black hole solution that has a non-vanishing TLN 
which clearly increases with mass and the tidal charge $q$. 

For higher values of $l$, from the expansion about infinity, we find the same solution for $2-P(u)$ but $Q(u)$ is $l$-dependent and reads
\be
Q(u) = -l (1+l) - 2l(1+l)u + \left(l^2 (q-4)+l (q-4)+2 (q-2)\right) u^2 + \ldots
\ee
The indicial equation reads
\be
\mathcal{R} (\mathcal{R} -1) - l (l+1)=0,
\ee
which has roots $\{ -l, 1+l \}$ with the particular solution being identical for all $l \geq 2$. 
Similar to the $l=2$ case, the solution associated with the indicial root of $-l$ contains the correct leading order tidal moment $r^l$. The solution associated with the other root of $1+l$ diverges at the horizon.

%\subsection{On Logarithmic TLNs}
%
%We will soon see that the asymptotic series expansion of braneworld solutions contain 
%terms of the form $\log (r/r_0), (\log (r/r_0) )^2$ where $r_0$ is a length parameter characterizing
%the undeformed solution and which generally sets the scale for the black hole horizon/wormhole throat. 
%
%
%\dotfill
%
%
%\dotfill
%
%
%\dotfill
%In the following, we review how the TLN appears as a tidal phase correction of the GW signal. 
%From Hinderer-Flanagan [ arXiv:0709.1915 ], 
%\bea
%\ddot{x}^i + \frac{M}{r^2}n^i &=& \frac{m_2}{2\mu} Q_{jk} \partial_i \partial_j \partial_k \frac{1}{r} - \frac{2}{5}x_j 
%\frac{d^5 Q^T_{ij}}{dt^5} ,\cr
%\ddot{Q}^n_{ij} + \omega^2_n Q^n_{ij} &=& m_2 \lambda_{1n} \omega^2_n \partial_i \partial_j \frac{1}{r} 
%-\frac{2}{5} \lambda_{1n} \omega^2_n \frac{d^5 Q^T_{ij}}{dt^5}. 
%\eea
%where 
%\be
%Q^T_{ij} = Q_{ij} + \mu r^2 (n_i n_j - \frac{1}{3} \delta_{ij} )
%\ee
%Above, $m_1,m_2$ are the masses of the binary with TLNs $\lambda_{1,2}$, $M,\mu$ are the total and reduced masses. For definiteness, star `1' is the one excited and $\omega_n, \lambda_n, Q^n_{ij}$ are the frequencies and contributions to TLN and $Q_{ij}$ from the $n$th radial mode (to be summed up). 
%
%In the following, we add in the logarithmic terms to $Q^T_{ij}$ and modify the first term of each RHS above by following the geodesic equation.
%
%\dotfill
%
%
%\dotfill
%
%
%\dotfill
%

%--------------------------------------------------------------------------------------
\section{Other examples of braneworld black holes and wormholes}
\label{Other}

We now proceed to consider other black hole and wormhole solutions with a similar computational
approach focussing on the $l=2$ case. For this work, we consider various black hole and wormhole solutions 
which are static and spherically symmetric.\footnote{See for example \cite{Dejan} for other braneworld black hole solutions.}
As we noted earlier,
in the generic case, there is no decoupling of $\delta \rho$ from the perturbation equations
and one has to solve a homogeneous third-order ODE for $H_0$. Nonetheless, the computational approach for studying the TLN is identical.
We begin by discussing some essential generic points before moving to studying specific black hole and wormhole geometries.

\subsection{Some preliminaries}
\label{Prelim}
Like in the case of the Tidal black hole solution, we do not know the analytic form for the metric perturbation $H_0 (r)$ for each 
family of black hole and wormhole geometries,
but will use both its near-horizon series expansion and that about the asymptotic infinity to deduce the series solution relevant 
for computing the TLN. 

For this work, the chosen geometries all enjoy points in their parameter spaces where they coincide with
certain limits of the Tidal black hole (see Table \ref{Table1}). For the Tidal black hole, we have identified the regular/divergent nature of 
each asymptotic series via the $q=0$ (Schwarzschild) limit which will again serve as the anchor limit for our analysis of two families of geometries: the 
Casadio-Fabbri-Mazzacurati (CFM) black holes and $\gamma$-wormholes (also found by the same authors). For a class of wormholes
found by Bronnikov and Kim, a point in its modulus space coincides with an analytic continuation $q=m^2$ of the Tidal black hole. Finally,
for a family of massless geometries found by Bronnikov-Melnikov-Dehnen, there is a choice of parameter for which it coincides with the massless limit of the 
Tidal black hole.

\begin{table}[h!]
\begin{center}
\begin{tabular}{ | m{3.6cm} | m{3.6cm} | m{3.9cm} | } 
 \hline
 Black hole/wormhole geometry & Relevant parameter domain  & Tidal black hole limits \\ 
\hline 
\hline
CFM black holes & $\beta \neq \frac{5}{4}$ & $\beta \rightarrow 1$ ($q=0$) \\ 
\hline
$\gamma-$ black holes & $\gamma >0$ & $\gamma \rightarrow 1$ ($q=0$) \\
\hline
Bronnikov-Kim wormholes & $r_0 >1, r_0 \neq 2$ & $r_0 \rightarrow 2$ ($q=m^2$) \\
\hline
Massless geometries & $C \in (-\infty, \infty )$ & $C \rightarrow 1$ ($m=0$)  \\
\hline
\hline
\end{tabular}
\caption{ In this Table, we summarize the relevant parameter domains for which we could compute the TLN based on our method, and the 
associated Tidal black hole limits for each family of black hole/wormhole geometry. See Table \ref{Table7} in Appendix \ref{AppB} for a more complete characterization in terms
of indicial roots for each geometry.}
\label{Table1}
\end{center}
\end{table}

Like the Tidal black hole, we also consider series solutions relevant for the `near-horizon' regime. The solutions that we study in this paper include wormholes and naked singularities apart from black hole geometries, and in those cases, we consider the expansion about the throat and singular loci.  By the `near-horizon regime', we generally refer to the small neighborhood of 
spherical surfaces along which the metric appears to be singular. Although the series solutions constructed in the near-horizon regime will not be useful for directly reading off the TLN, 
obtaining their general form by computing the indicial roots serves as a necessary condition for their
regularity at the horizon or throat as we have seen in the case of the Tidal black hole. 

Expanding about
a metric singularity in each case (for us, this is either a
black hole horizon or wormhole throat), we seek the near-horizon series solution for each family of geometries,
after ensuring that the metric singularity is a regular point for the Frobenius method. Let the expansion parameter be $x= \frac{r-r_h}{L}$ where $r_h$ could be either the black hole throat or the wormhole throat and $L$
is some length parameter of each family of solutions. We will also be treating a couple of cases of naked singularities, in which
case, $r_h$ is simply the singular surface. In all cases, $r=r_h$ describes a codimension-two metric singularity which may or may not
cloak physical curvature singularities. Writing \eqref{ThirdOrderODE} in the form 
\be
H'''_0 + \frac{\mathcal{P}_1 (x)}{x}H''_0 + \frac{\mathcal{P}_2 (x)}{x^2}H'_0 + \frac{\mathcal{P}_1 (x)}{x^3} H_0 = 0,
\ee
we compute the following limits from the metric
\be
\lim_{x\rightarrow 0} \mathcal{P}_1 \equiv \mathcal{P}_{10},\,\,
\lim_{x\rightarrow 0} \mathcal{P}_2 \equiv \mathcal{P}_{20},\,\,
\lim_{x\rightarrow 0} \mathcal{P}_3 \equiv \mathcal{P}_{30},\,\,
\ee
from which we compute the indicial roots or the roots of the cubic equation
\be
\mathcal{R} (\mathcal{R} -1)(\mathcal{R} -2) + \mathcal{R} (\mathcal{R} -1)\mathcal{P}_{30} + 
\mathcal{R} \mathcal{P}_{20} + \mathcal{P}_{10} = 0.
\ee

We find that typically, 
there are points in each parameter space which correspond to singularities in the near-horizon series solutions. 
These `exotic' points are marked by discontinuities in the first-derivative of the horizon radius as a function of the parameter, and where
indicial roots separate into different branches (see Table \ref{Table7} of Appendix \ref{AppB}). We will not rigorously study these cases here, 
but will briefly comment on them in Appendix ~\ref{AppB}. Since we don't have an analytic solution to determine regularity at horizon directly, 
we rely on smooth limits to the Tidal black hole for this purpose. Each regular series solution (about the asymptotic infinity) that we construct
for computing the TLN is smoothly connected to various limits of  $y_{\{r,d,q \}} (u,q)$ (see Table \ref{Table1}). There are domains of parameters where
the indicial roots (for near-horizon expansion) are all non-negative, indicating that all independent solutions are regular at horizon. 
For our work, we will nonetheless restrict ourselves to regular solutions which have a clear GR limit, and hence the final series solution 
we choose to compute the TLN is always that associated with (limits of) $y_r (u,q)$. 

There are also domains of parameter spaces where one of the 
indicial roots is negative. 
To explicitly confirm that in the relevant limit ($q\rightarrow 0$ or $m\rightarrow 0$), they are not associated with the asymptotic series solution we use to compute the TLN, we work out the series solution belonging to the negative root in the cases that it arises. All such cases have the set $\{-1,0,1\}$ for their indicial roots in the near-horizon expansion, apart from those
not smoothly connected to any limits of the Tidal black hole.

For expanding about $r=\infty$, we define $u=L/r$ where $L$ is some length parameter in the undeformed solution, and recast the ODE in \eqref{ThirdOrderODE} into the following form
\be
\label{thirdHomU}
H'''_0 + \frac{1}{u} \mathcal{P}_1 H''_0 + \frac{1}{u^2} \mathcal{P}_2 H'_0 
+ \frac{1}{u^3} \mathcal{P}_3 H_0 = 0,
\ee
where 
\be
\mathcal{P}_1 = 6 - \frac{L C_2}{C_3 u}, \,\,\, 
\mathcal{P}_2 = 6 - \frac{2L C_2}{C_3 u} + \frac{L^2 C_1}{C_3 u^2},\,\,\,
\mathcal{P}_3 = - \frac{L^3 C_0}{C_3u^3}.
\ee
The indicial equation reads 
\be
\label{IndicialHor}
\mathcal{R} (\mathcal{R}-1)(\mathcal{R}-2) + \mathcal{R}(\mathcal{R}-1) \mathcal{P}_{10} + \mathcal{R} \mathcal{P}_{20} + \mathcal{P}_{30} = 0
\ee
of which roots $\mathcal{R}_1 > \mathcal{R}_2 >\mathcal{R}_3$ determine the form of the general solution. 
In contrast to the near-horizon expansion for
all the solutions considered in this work, we find the universal values
\bea
&&\mathcal{P}_{10} =2,\,\, \mathcal{P}_{20} =-6,\,\, \mathcal{P}_{30} =0, \cr
&& R_1 = 3, \,\, R_2 = 0, \,\, R_3 = -2. 
\eea
This leads to the general solution being the linear combination of
\bea
S_1 (u) &=& u^{3} (1 + a_1 u + a_2 u^2 + a_3 u^3 + \ldots),\cr
S_2 (u) &=& V_1 S_1 (u) \log (u) + \left( 1 + b_1 u + b_2 u^2 +  \ldots \right),\,\,
\cr
S_3 (u)&=& \frac{1}{u^2} \left( 1 + d_1 u + d_2 u^2 + \ldots + d_5 u^5 + \ldots \right)
+ W_2 \log (u) \left( c_0 + c_1 u + c_2 u^2   + \ldots \right) + W_3 (\log (u))^2 S_1 (u). \nonumber \\
\eea
As mentioned earlier,
each family of solutions has  a certain limit within its moduli space which 
reduces uniquely to the Schwarzschild solution or certain limits of the regular tidal-deformed metric
of the Tidal black hole. This yields consistency checks for regularity of each solution 
at the horizon or wormhole throat whenever there are negative indicial roots in the near-horizon expansion.
In all cases, we take $S_3 (u)$ 
to be the series relevant for picking the TLN, as it is the one associated with $y_r (u,q)$. We also need the full metric component $h_{tt}$ in each case which is
collected separately in the Appendix \ref{AppB}.

For higher values of $l$,
in all other braneworld solutions we consider in this work, we find the following universal values for the roots of the indicial equation: 
\bea
&&\mathcal{P}_{10} = 2, \,\, 
\mathcal{P}_{20} = -l (1+l), \,\,
\mathcal{P}_{30}=0, \cr
&& \mathcal{R}_1 = l+1, \,\, \mathcal{R}_2 =0,\,\, \mathcal{R}_3 = -l.
\eea
Similar to the specific case of $l=2$, this leads to the general solution being the linear combination of 
\bea
S_1 (u) &=& u^{l+1} (1 + a_1 u + a_2 u^2 + a_3 u^3 + \ldots),\cr
S_2 (u) &=& V_1 S_1 (u) \log (u) + \left( 1 + b_1 u + b_2 u^2 +  \ldots \right),\,\,
\cr
S_3 (u)&=& \frac{1}{u^l} \left( 1 + d_1 u + d_2 u^2 + \ldots + d_5 u^5 + \ldots \right)
+ W_2 \log (u) \left( 1 + c_1 u + c_2 u^2   + \ldots \right) + W_3 (\log (u))^2 S_1 (u). \nonumber \\
\eea
In the following, we will compute the TLN for various black hole and wormhole geometries, first by working out
the asymptotic series solutions, identifying their Tidal black hole counterparts before presenting details concerning their
near-horizon expansions and regular/divergent behavior at horizon. As mentioned, the indicial roots for the asymptotic expansion about infinity take on the universal values $\{ 3,0,-2 \}$ and we denote their corresponding series solutions by $S_1 (u) , S_2 (u) , S_3 (u)$ respectively, with $S_3 (u)$ being the one relevant for determining the TLN. The indicial roots for the near-horizon expansion exhibit a much richer branch structure that bifurcates at points where the horizon radius - as a function of some parameter - is discontinuous in its first/second derivatives (kinks). For each family of solutions, there are regions in the parameter space where all indicial roots are positive and hence all independent series solutions are regular at horizon. For regions which share the same identical roots $\{1,0,-1 \}$ as the Tidal black hole, we denote their corresponding series solutions by $\tilde{S}_r (x), \tilde{S}_p (x), \tilde{S}_d (x)$ respectively. Finally, for regions that are not smoothly connected to some limit of the Tidal black hole, we briefly comment on them in a separate Appendix which contains a summary of all the indicial roots.

\begin{table}[h!]
\begin{center}
\begin{tabular}{   | m{2.5cm} | m{2.0cm} || m{2.5cm} | m{2.0cm}|  } 
 \hline
Series solution  (infinity)  & Indicial Roots & Series solution (near-horizon) & Indicial Roots \\ 
\hline 
$S_1 (u)$ & 3 & $\tilde{S}_d (x)$ & -1\\
\hline
$S_2 (u)$ & 0 & $\tilde{S}_p (x)$ & 0\\
\hline
$S_3 (u)$ & -2 & $\tilde{S}_r (x)$ & 1\\
\hline
\end{tabular}
\caption{For the sake of reading clarity, we tabulate the nomenclature for various series solutions computed for each class of black hole and wormhole geometries in Sections ~\ref{CFM} - ~\ref{Massless}.   }
\label{Table2}
\end{center}
\end{table}

\subsection{CFM black holes}
\label{CFM}

This family of black hole solutions was found by Casadio, Fabbri and Mazzacurati in \cite{Casadio}. 
The metric components read
 $$f(r) = 1- \frac{2m}{r}, \,\, g(r) = \frac{1-\frac{3m}{2r}}{\left( 1- \frac{2m}{r} \right)\left( 1 - \frac{m}{2r}(4\beta-1)    \right)}.$$ 
The horizon is located at $r=r_h = 2m$
and the limit $\beta \rightarrow 1$ corresponds to the Schwarzschild ansatz. 

\begin{figure}[h!]
\centering
\includegraphics[scale=0.61]{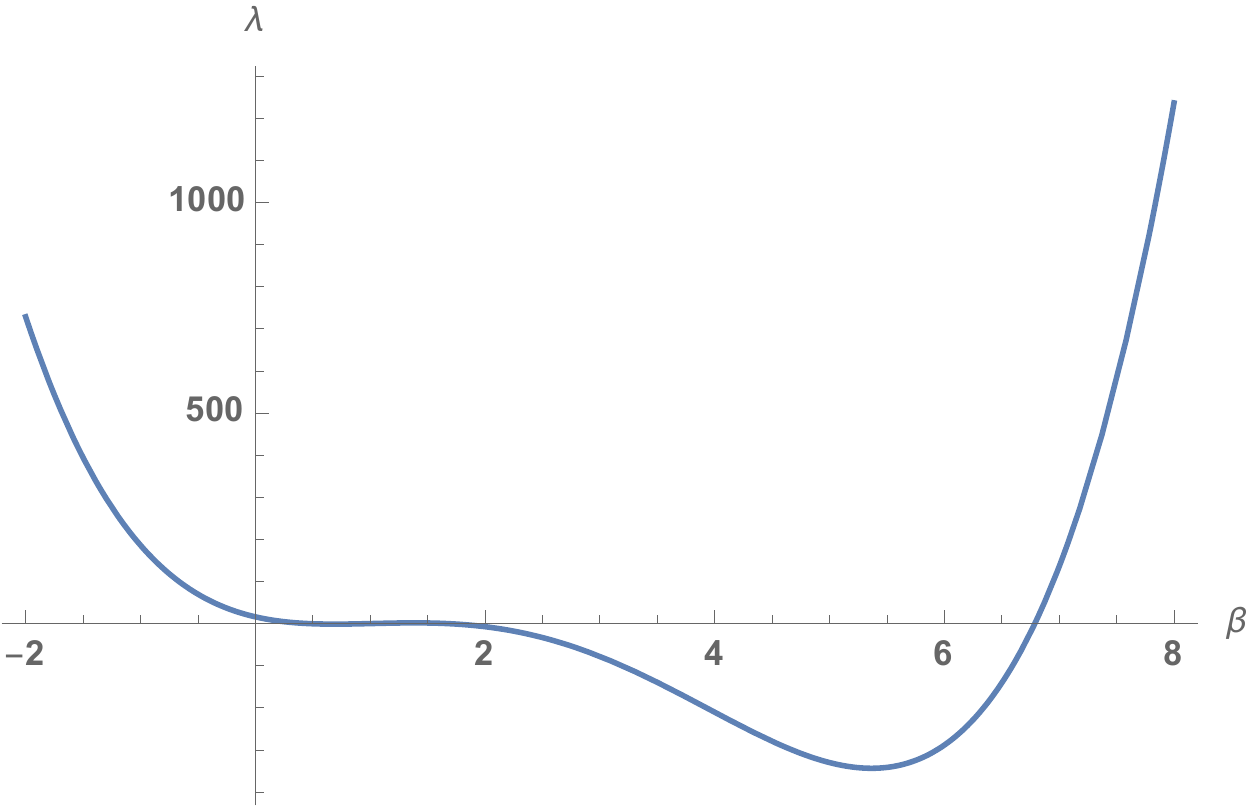}
\caption{Plot of the TLN $\lambda$ (in units of $m^5$ ) vs the parameter $\beta$. The Schwarzschild limit corresponds to $\beta=1$ and we have symmetric traversable wormhole geometries for $\beta \geq \frac{5}{4} $ although there is no transitional behavior of $\lambda$ near this critical value. The other zeroes do not correspond to any distinct causal structure of the solution.
In Figure \ref{CFM2}, we zoom in onto the neighborhood of $\beta=1$. }
\label{CFM1}
\end{figure}

\begin{figure}[h!]
\centering
\includegraphics[scale=0.71]{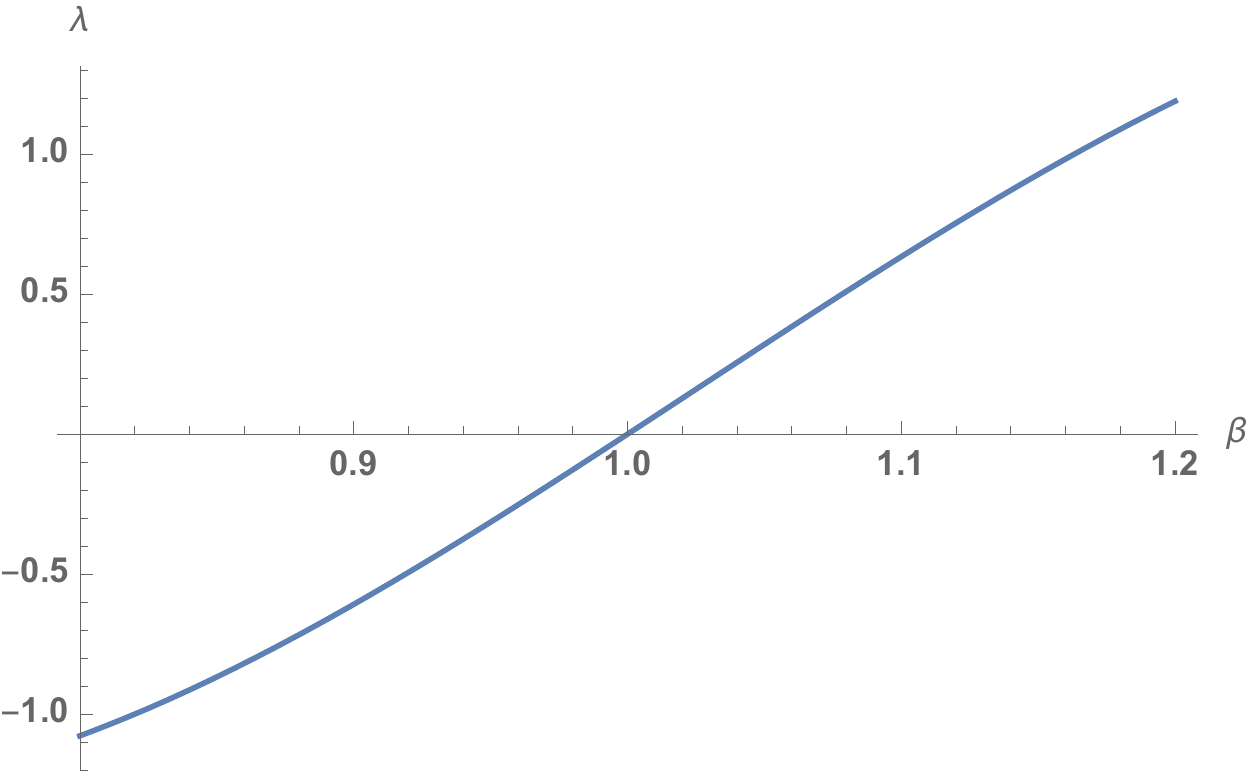}
\caption{Plot of the TLN $\lambda$ (in units of $m^5$ ) vs the parameter $\beta$ near $\beta \approx 1$. Near this transition value, solutions with $\beta >1$ are non-singular in nature while those with $\beta<1$ contain curvature singularities.  }
\label{CFM2}
\end{figure}

We first present the asymptotic series solutions.
For $S_1 (u)$, we find 
\be
S_1 (u) = u^3 +\frac{1}{4} (13 \beta - 1)  u^4 +
 \frac{1}{70} \left(533 \beta^2-112 \beta+79\right) u^5 + \ldots
\ee
which, in the limit $\beta \rightarrow 1$, becomes
\be
S_1 (u) \rightarrow u^3 (1 + 3u + \frac{50}{7} u^2 + \ldots ) = y_d (u,q=0).
\ee
This is precisely the series that we identify to be divergent at the horizon. The $S_2 (u), S_3 (u)$ series read
\bea
\label{CFMy2}
S_2 (u) &=& -(5 - 8 \beta + 3 \beta^2)   y_1(u) \log (u) + 1+2 u -\frac{1}{2} (3 \beta -11) u^2  + \ldots   \cr
\label{CFMy3}
S_3 (u) &=& \frac{1}{4} (\beta-1)^2 \left(33 \beta^2-64 \beta+15\right) y_1 (u) \log^2 (u) \cr
&&\qquad + \log (u) \left(
\frac{1}{2}(1-\beta)(11\beta-3) (1+ 2u - \frac{1}{2} (3 \beta -11) u^2  + \ldots )
\right) \cr
&&\qquad + \frac{1}{u^2} (1 -\frac{1}{2} (11 \beta - 7) u + \frac{1}{12} \left(55 \beta^3-180 \beta^2+158 \beta-33\right) u^3 + \ldots 
\eea
In the $\beta \rightarrow 1 $ limit, the $\log (u)$ series in $S_2(u)$ vanishes and we find
\be
\label{y2limitcfm}
S_2 (u) \rightarrow  y_p (u,q=0)/\alpha
\ee
which is precisely the particular solution in the Schwarzschild limit.\footnote{We should note that in solving for the series $S_2 (u)$, there is an arbitrary constant $b_3$ which parametrizes $S_1 (u)$ and setting it to vanish implies that 
in the $q \rightarrow 0$ limit, we have \eqref{y2limitcfm}.}
In the case of the Tidal black hole, the $\delta \rho$ equation decouples and we have an arbitrary constant characterizing the arbitrary strength of the quadrupole moment. But this is not induced by some tidal moment (there is no accompanying $1/u^2$). The cofficient for $u^3$ in $S_2 (u)$ is not what we seek in the traditional TLN definition. We should focus on the solution branch with $1/u^2$ representing the presence of the tidal moment due to an external body. Now in the same limit, both $\log (u)$ and $( \log (u) )^2$ terms in $S_3 (u)$ vanish and we have
\be
\label{y3limitcfm}
S_3 (u) \rightarrow \frac{1}{u^2} - \frac{2}{u} = y_r (u,q=0)
\ee
which is the $q=0$ limit of the regular tidal-deformed Schwarzschild solution which has vanishing TLN in the absence of
any $r^3$ term.\footnote{ Again, we should note that \eqref{y3limitcfm} is obtained by setting $d_2 = d_5=0$ and these parameters are otherwise arbitrary. }
From $h_{tt}$ (see Appendix \ref{AppA3}), 
we read off the TLN to be (see Figures \ref{CFM1} and \ref{CFM2})
\be
\lambda =m^5  \frac{1}{72} (\beta-1) \left(220 \beta^3-1963 \beta^2+3358 \beta-1155\right).
\ee
Phenomenological interest lies in the small neighborhood of the critical value $\beta =1$ where for solutions with $\beta >1$, the TLN is positive whereas it is negative for $\beta <1$.  
The $\beta <1$ solutions are black holes that are Schwarzschild-like in nature with an event horizon at $r_h = 2m$ , whereas
the $\beta >1$ solutions are non-singular wormhole geometries. For $1 < \beta < \frac{5}{4}$, their Carter-Penrose diagrams
resemble the form of the Kerr black hole, and for $\beta \geq \frac{5}{4}$, we have symmetric traversable wormholes. 
In this family of solutions, near the critical point, the sign of the TLN for solutions indicates the presence (-) or absence (+)
of the black hole curvature singularity. 

%---------------------------------------------------
For the expansion
about the horizon $r=r_h$, we note that as a function of $\beta$,
$r_h$ has a discontinuity in second-derivative at $\beta= \frac{5}{4}$, 
beyond which it increases with $\beta$. In Table \ref{Table3}, we summarize
the physical interpretations of the geometry and the horizon for different $\beta$ intervals. 
The indicial roots separate into two branches following this distinguished value,
being the set $\{ -1,0,1 \}$ for $\beta < \frac{5}{4}$ and $\{0, \frac{1}{2},1 \}$
for $\beta > \frac{5}{4}$. The Schwarzschild limit lies at $\beta =1$. 

For $\beta> \frac{5}{4}$, the indicial roots are all positive so all the solutions of the ODE are regular at the horizon.
For $\beta < \frac{5}{4}$, the solutions are smoothly connected to the Schwarzschild $q=0$ limit, 
and their indicial roots are identical. We explicitly check
that for the negative root, the $q=0$ limit yields
\be
\lim_{q=0} \tilde{S}_d (x) = \tilde{S}_d (x,q=0) + 3\tilde{S}_p (x,q=0),
\ee
and thus is independent of the near-horizon solution $S_r (x;q=0)$ corresponding to $S_3 (u)$. We also find that 
\be
\lim_{q=0} \tilde{S}_r (x) = \tilde{S}_r (x,q=0), \,\,\, 
\lim_{q=0} \tilde{S}_p (x) = \tilde{S}_p (x,q=0).
\ee
Our method doesn't rigorously apply for the isolated case of $\beta =\frac{5}{4}$ at which the near-horizon series expansion develops a singularity
paralled by the kink in $r_h$ as a function of $\beta$. In this case, there is nonetheless a positive indicial root and the asymptotic series $S_3 (u)$ has a smooth limit 
at this point, so it is nevertheless possible that the solution of which asymptotic expansion is $S_3 (u)$ is regular at the horizon.

\begin{table}[h!]
\begin{center}
\begin{tabular}{  |c|c| } 
 \hline
$r_h$  & Indicial Roots \\ 
\hline 
 Sch-BH, $\beta <1, r_h = 2m$ & $\{-1,0,1\}$ \\
 Kerr-WH, $\beta  \in (1,\frac{5}{4} ), r_h = 2m$ &  $\{-1,0,1\}$ \\ 
 RN-WH, $\beta  = \frac{5}{4}, r_h = 2m$  &  $\{-1,-\frac{1}{4}(1 \pm \sqrt{9+4l+4l^2} )\}$ \\ 
 WH, $\beta > \frac{5}{4} , r_{h} = \frac{m}{2}(4\beta -1 )$ &   $\{0,\frac{1}{2},1\}$ \\ 
\hline
\end{tabular}
\caption{In the above, we use the abbreviations  WH (wormholes), Sch-BH (Schwarzschild black holes), Kerr-WH refers to completely regular wormhole geometries of which CP diagram resembles the form of that of Kerr and RN-WH refers to spacetimes with the casual structure of extremal 
Reissner-Nordstr$\ddot{\text{o}}$m with a horizon cloaking a time-like singularity.  }
\label{Table3}
\end{center}
\end{table}

%---------------------------------------------------

%----------------------------------------------------------------------
\subsection{$\gamma$-wormholes}
\label{Gamma}

This family of solutions was studied by Casadio, Fabbri and Mazzacurati in \cite{Casadio}, and contains both 
pathological naked singularities as well as regular wormhole geometries. The metric components read
\be
f(r)
=  \frac{1}{\gamma^2} \left( \gamma - 1 + \sqrt{1- \frac{2\gamma m}{r}} \right)^2, \,\,\,
g(r) = \left( 1 - \frac{2\gamma m}{r} \right)^{-1},\,\,
\ee
with $\gamma =1$ being the Schwarzschild limit.  
For $\gamma <1$, the metric is singular at  
\be
r_{s} = \begin{cases}
\frac{2m}{2-\gamma} \equiv r_h  \\
2m \gamma \equiv r_0, 
\end{cases}
\ee
where $r_h = \frac{2m}{2-\gamma} $ is a null and singular surface along which the Ricci scalar $R$
diverges as $R\sim 
1/ \left(   
\sqrt{r-r_0} - \sqrt{r_h - r_0}
\right)$. 
For $\gamma >1$, the only metric singularity lies 
at $r_0=2m \gamma$. There is a turning/minimum point (for all timelike geodesics) at $r=2m\gamma$ where 
all curvature invariants are regular. The causal interpretation is that of a wormhole solution as explained in \cite{Casadio}. 

We first present the asymptotic series solutions.  Corresponding to the highest root of the indicial equation, 
\be
S_1 (u) = 
u^3+\frac{1}{4} (13 \gamma -1)    u^4+ \frac{1}{70} \left(533 \gamma ^2-42 \gamma +9\right) u^5+\mathcal{O} ( u^6 ),
\ee
We find that
in the limit $q\rightarrow 1$, this reduces precisely to the divergent branch of the tidal-deformed Schwarzschild solution, i.e.
$$
\lim_{\gamma \rightarrow 1} S_1 (u) = y_d (u,q=0)
$$
Thus, we discard this solution branch for ensuring regularity at horizon. We also find the following solution 
\bea
S_2 (u) &=& \frac{1}{5} \left(-11 \gamma ^2+34 \gamma -23\right) y_1 (u) \log (u) + 1 + 2 u + \frac{9-\gamma }{2} u^2 + \ldots,
\eea
We note that in solving for the series $S_2 (u)$, there is an arbitrary constant $b_3$ which 
parametrizes $S_1 (u)$ and setting it to vanish implies that 
in the $q \rightarrow 1$ limit, this solution reduces exactly to the particular solution of the tidal-deformed Schwarzschild solution, i.e.
$$
\lim_{\gamma \rightarrow 1} S_2 (u) = y_p (u,q=0).
$$
Finally, corresponding to the only negative root of the indicial equation, we have
\bea
S_3 (u) &=& \frac{1}{20} (\gamma -1)^2 \left(121 \gamma ^2-286 \gamma +69\right) y_1 (u) \log^2 (u)  \cr
&& \qquad  \frac{1}{2} (-11 \gamma ^2+14 \gamma -3) \log (u) \left(1 + 2 u + \frac{9-\gamma }{2} u^2 + \ldots \right) \cr
&& \qquad \frac{1}{u^2} \left( 1 + \frac{1}{2} (7-11 \gamma ) u  + \frac{1}{12} \left(55 \gamma ^3-196 \gamma ^2+189 \gamma -48\right) u^3 + \ldots \right) 
\eea
where we found that setting the arbitrary constants $d_2=d_5=0$ allow us to identify $S_3 (u)$ with the regular solution of the
tidal-deformed Schwarzschild solution, i.e. 
$$
\lim_{\gamma \rightarrow 1} S_3 (u) = y_r (u,q=0).
$$
Thus, just as in the case of the CFM black holes, we found that the three series solutions can be parametrically connected to their corresponding series solutions of the $q=0$ limit of the tidal-deformed Tidal black hole in a unique way, and that the one corresponding to the negative root of the indicial equation $S_3 (u)$ is the one
relevant for computing TLN. 

\begin{figure}[h!]
\centering
\includegraphics[scale=0.6]{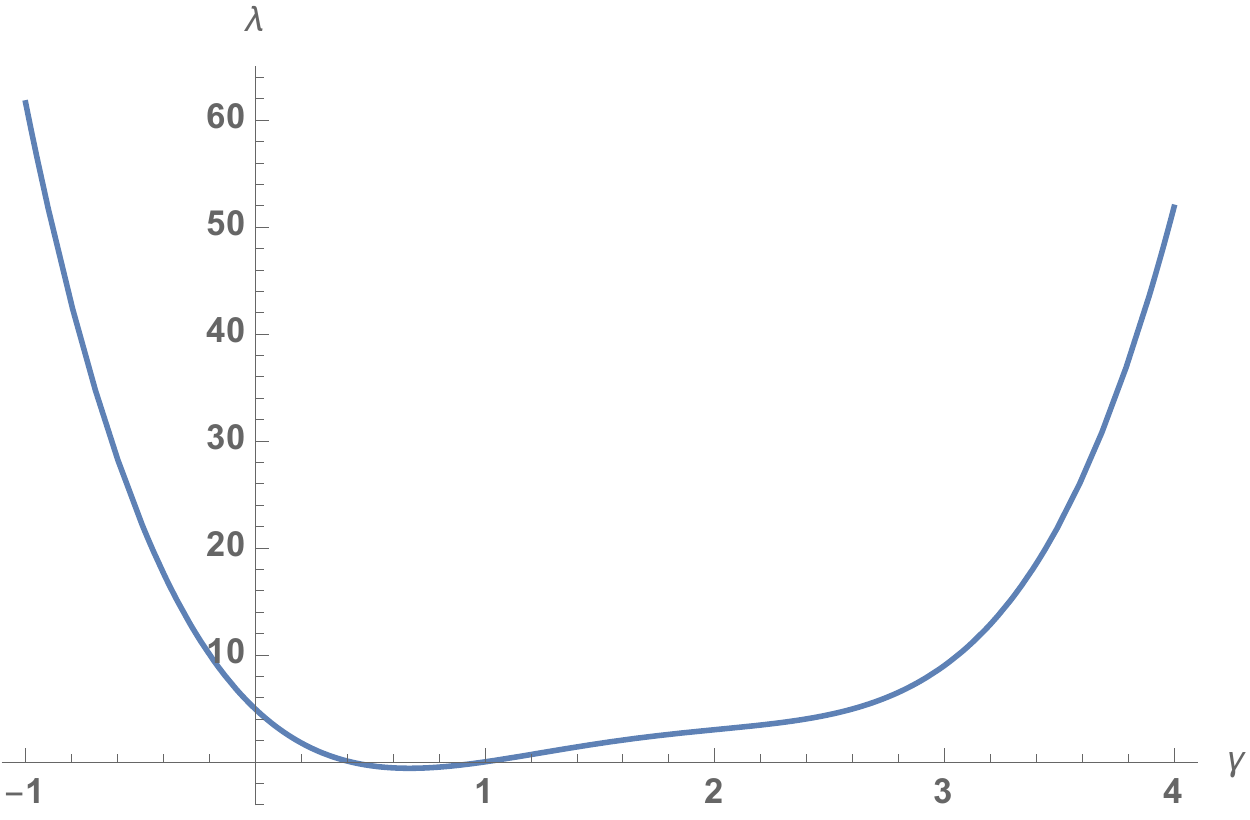}
\caption{Plot of $\lambda$ (in units of $m^5$) vs $\gamma$ for the family of  $\gamma$-wormhole solutions. The solutions contain naked singularities for
$\gamma <1$ whereas we have regular wormhole geometries for $\gamma >1$.  }
\label{gamma1}
\end{figure}

\begin{figure}[h!]
\centering
\includegraphics[scale=0.6]{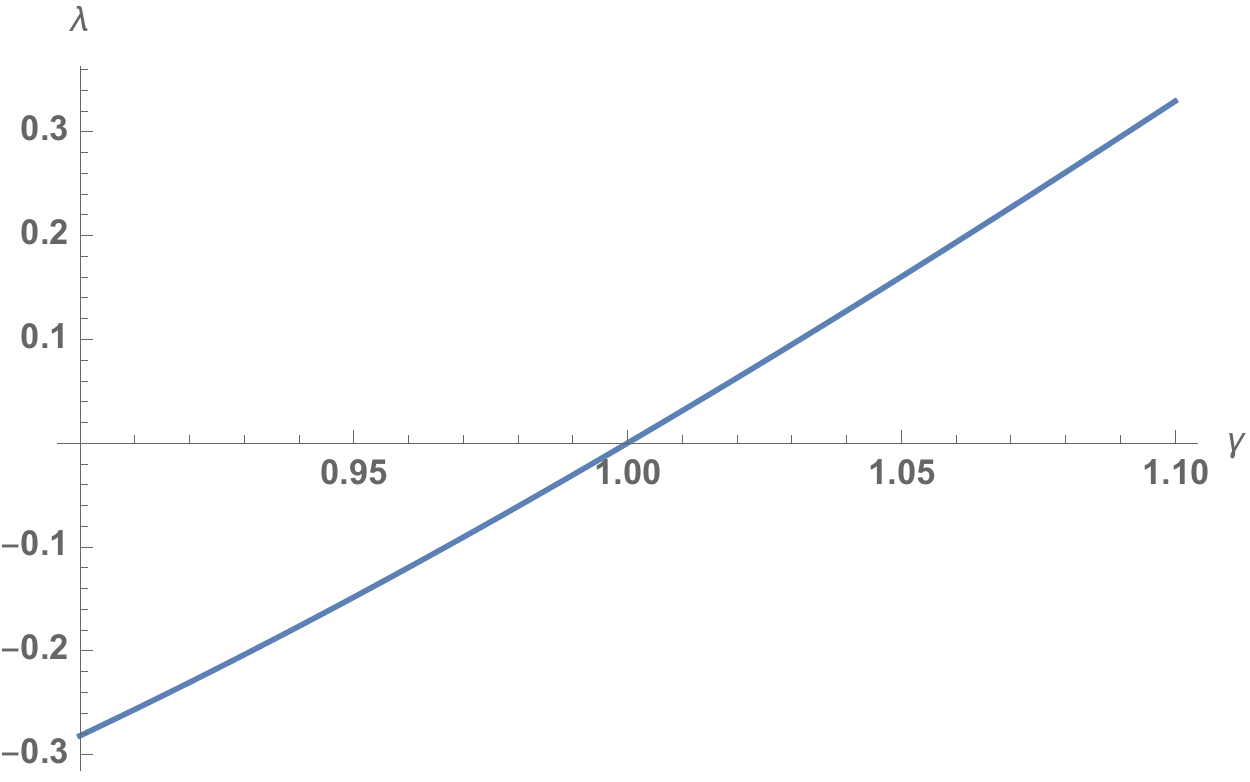}
\caption{Plot of $\lambda$ (in units of $m^5$) vs $\gamma$ near Schwarzschild limit $\gamma = 1$, showing clearly 
that in the vicinity of the transition point, the sign of $\gamma$ corresponds to the presence (-) / absence (+) of 
curvature singularities.    }
\label{gamma2}
\end{figure}

Taking into account the background metric, from the series expansion of $h_{tt} = f(r) S_3 (u)$ 
we can read off the TLN to be (see Figures \ref{gamma1} and \ref{gamma2} )
\be
\lambda = m^5 \frac{1}{144} (\gamma -1) \left(233 \gamma ^3-1284 \gamma ^2+2232 \gamma -711\right).
\ee
In the neighborhood of the critical value $\gamma=1$, we find that $\lambda > 0$ for $\gamma >1$ which corresponds to 
non-singular wormhole geometries whereas $\lambda <0$ for $\gamma<1$ which corresponds to naked singularities. 
This is similar to the scenario in our study of tidal-deformed CFM black hole solutions where a positive $\lambda$ labels the non-singular
wormhole branch of the solution class whereas negative $\lambda$ pertains to Schwarzschild-type black holes near the critical Schwarzschild point. In both cases,
$\lambda$ is a degree-four polynomial in the parameter with one of the polynomial roots associated with the Schwarzschild limit. 

For the near-horizon series expansions, we find that the indicial roots separate into three branches as
shown in Table \ref{Table4} below. Apart from the case of $\gamma =1$ which coincides with the $q=0$ Schwarzschild limit, 
all indicial roots are positive and hence the solutions are all regular at the horizon.

\begin{table}[h!]
\begin{center}
\begin{tabular}{  |c|c| } 
 \hline
$r_h$  & Indicial Roots \\ 
\hline 
 NS, $\gamma <1, r_h = \frac{2m}{2-\gamma}$ & $\{0,1,2\}$ \\
 Sch-BH, $\gamma = 1, r_h = 2m$ &  $\{-1,0,1\}$ \\ 
 WH, $\gamma > 1 , r_h = 2m\gamma$ &   $\{0,\frac{1}{2},1\}$ \\ 
\hline
\end{tabular}
\caption{In the above, NS refers to Naked Singularities for which $r_h$ is a null surface where the Ricci scalar diverges. }
\label{Table4}
\end{center}
\end{table}

\subsection{Bronnikov-Kim wormholes}
\label{BK}

In \cite{Bronnikov}, Bronnikov, Melnikov and Dehnen presented a powerful solution-generating technique
to construct exact braneworld black hole and wormhole solutions for Randall-Sundrum theory considered 
in our work, covering the previous two families of solutions. In this and the subsequent sections, 
we study the two concrete examples mentioned in their seminal work. 

We first study example 3 of \cite{Bronnikov} where the line element reads
$$
f(r) = \left( 1 - \frac{2\tilde{m}}{r} \right)^2,\,\, 
g(r) = \left( 1 - \frac{r_0}{r} \right)^{-1} \left( 1 - \frac{r_1}{r} \right)^{-1}, \,\,\, r_1 = \frac{\tilde{m}r_0}{r_0-\tilde{m}},
$$
with the parameter $r_0$ determining the causal structure as follows:
\begin{itemize}
\item $r_0 < \tilde{m}$: naked singularity at $r=2\tilde{m}$,
\item $\tilde{m}<r_0<2\tilde{m}$: wormhole with throat at $ r_1>2\tilde{m}$,
\item $r_0 = 2\tilde{m}$: extremal Reissner-Nordstr$\ddot{\text{o}}$m,
\item $r_0 > 2\tilde{m}$: wormhole with throat at $r_0$.
\end{itemize}
Unlike the previous two cases, for this solution,
there is no natural limiting procedure to send it to Schwarzschild, but to the $q=m^2, m= 2 \tilde{m}$ limit of the Tidal black hole or
equivalently the extremal
Reissner-Nordstr$\ddot{\text{o}}$m (in ordinary GR) at some finite value of the radial
coordinate $r_0 = 2\tilde{m}$. Since the tidal charge $q$ is originally negative, this is an analytic continuation.

\begin{figure}[h!]
\centering
\includegraphics[scale=0.75]{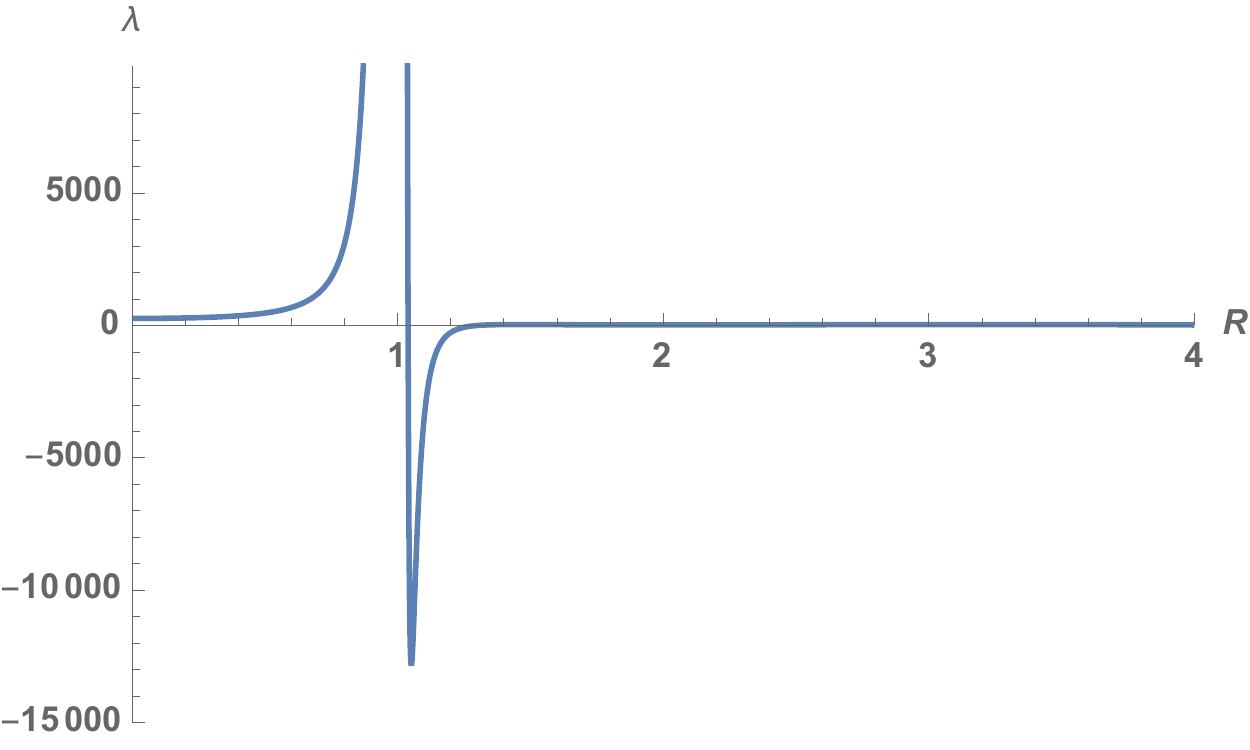}
\caption{Plot of $\lambda$ (in units of $\tilde{m}^5$) vs $R$. We note that $R=1$  marks the transition to a naked singularity with a vertical asymptote, the left of which pertains to naked singularities and the right of which is associated with regular wormhole geometries. At $R=2$, we have the extremal Reissner-Nordstr$\ddot{\text{o}}$m metric. }
\label{N1}
\end{figure}

\begin{figure}[h!]
\centering
\includegraphics[scale=0.65]{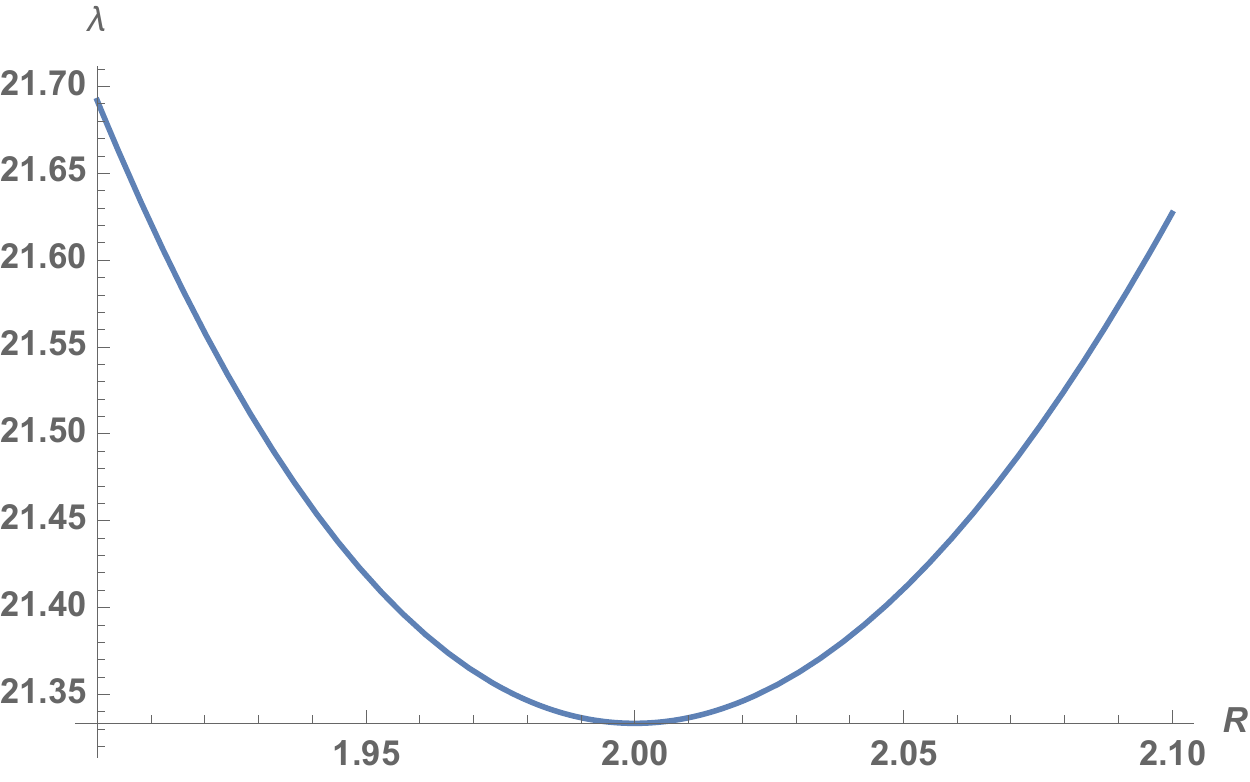}
\caption{Plot of $\lambda$ (in units of $\tilde{m}^5$) vs $R$ near the local minimum point $R=2$ limit which pertains to the extremal Reissner- Nordstr$\ddot{\text{o}}$m metric. This is a solution which is also part of the Tidal black hole family, albeit an extremal solution with positive tidal charge $q=m^2$. The value of $\lambda$ at $R=2$ was checked to agree with \eqref{lambdaTidal} in such a limit.  }
\label{N2}
\end{figure}

We find the following series solutions (taking the expansion variable to be $u = \tilde{m}/r$, and defining $R=r_0/\tilde{m}$)
\bea
S_1 (u) &=& u^3 + \frac{13 R^2-4 R+4}{8 (R-1)} u^4 + \frac{533 R^4-448 R^3+592 R^2-288 R+144}{280 (R-1)^2}  u^5 + \ldots  \cr
S_2 (u) &=&  -\frac{(R-2)^2 \left(15 R^2-92 R+92\right)}{10 (R-1)^2} y_1 (u) \log (u) + 1 + 4 u + \frac{3 R^2-36 R+36}{2-2 R} u^2 + \ldots \cr
S_3 (u) &=& \frac{(R-2)^4 \left(165 R^4-1192 R^3+2296 R^2-2208 R+1104\right)}{160 (R-1)^4} y_1 (u) \log^2 (u) \cr
&& \,\, -(((-2 + R)^2 (12 - 12 R + 11 R^2))/(8 (-1 + R)^2)) \log (u) ( 1 + 4 u +  \frac{3 R^2-36 R+36}{2-2 R}   u^2 + \ldots ) \cr
&&\,\, +\frac{1}{u^2} (1 + \frac{11 R^2-28 R+28}{4-4 R} u +   \frac{8}{3}u^2 + \ldots ) \nonumber \\
\eea
where for $S_2 (u)$ and $S_3(u)$, we have picked uniquely appropriate values of $b_3=0, d_2=\frac{8}{3}, d_5=0$ such that 
in the $R=2$ limit, each reduces to a corresponding series solution in the $q=m^2 = 4\tilde{m}^2$ limit
of the Tidal black hole. 
Specifically, we find in this extremal limit
\bea
\lim_{R\rightarrow 2} S_1 (u) &=&  y_d (u,q=m^2), \cr
\lim_{R\rightarrow 2} S_2 (u) &=&  y_p (u,q=m^2) / \alpha, \cr
\lim_{R\rightarrow 2} S_3 (u) &=& y_r (u,q=m^2),
\eea
with the last series solution being the regular solution from which one determines the TLN. Taking into account
the background metric, from the series expansion of 
$h_{tt} = f(r)S_3 (u) $, 
we read off the TLN to be  (see Figures \ref{N1} and \ref{N2} )
\be
\lambda =\tilde{m}^5 \frac{55 R^8-2073 R^7+19389 R^6-83064 R^5+200948 R^4-298640 R^3+278448 R^2-153344 R+38336}{144 (R-1)^4}.
\ee
We note that the vertical asymptote at $R=1$ corresponds to the limit in which the wormhole geometry degenerates into a naked 
singularity. Another notable point is at $R=2$ which corresponds to the extremal Reissner-Nordstr$\ddot{\text{o}}$m solution. This turns out
to be a local minimum point, with $$\lim_{R\rightarrow 2} \lambda = \frac{64}{5}\tilde{m}^5 = \frac{2}{3}m^5,$$ in agreement
with the $q=m^2$ limit of \eqref{Yr}. Unlike the extremal black hole solution which has a time-like curvature singularity behind a horizon, the wormhole geometries are globally regular.  

For the near-horizon series expansions, we find that the indicial roots are separated into several branches. 
They are all positive for $r_0 >1, r_0 \neq 2$ where we can trust their regularity at the horizon and thus the computation of TLN.
At the distinguished point
$R=2$ where it coincides with the $q=m^2$ analytic continuation of the Tidal black hole, the indicial roots take 
on a different set of values however, and the series solutions are not smoothly connected to
those of the Tidal black hole. There is a vertical asymptote at $R=1$ and for $R<1$, the indicial roots are $\{-2,0,2 \}$, 
and thus these series solutions cannot be smoothly deformed to any limit of the Tidal black hole.

\begin{table}[h!]
\begin{center}
\begin{tabular}{  |c|c| } 
 \hline
$r_h$  & Indicial Roots \\ 
\hline 
 NS, $r_0 <\tilde{m}, r_s = 2\tilde{m} $ & $\{-2,0,2\}$\\ 
WH, $r_0 \in [\tilde{m}, 2\tilde{m}) , r_{h} = r_1$  & $\{0,\frac{1}{2},1\}$\\ 
WH, $r_0 > 2\tilde{m} , r_{h} = r_0 $  & $\{0,\frac{1}{2},1\}$\\ 
RN, $r_0 =  2\tilde{m} , r_{h} = r_0  $ & $\{ -2,
-\frac{1}{2}(1 \pm \sqrt{9+4l+4l^2} )\}$\\ 
\hline
\end{tabular}
\caption{In the above, RN refers to the extremal Reissner-Nordstr$\ddot{\text{o}}$m which is the $q=m^2$ analytic continuation of the Tidal black hole. }
\label{Table5}
\end{center}
\end{table}

\subsection{Massless Geometries}
\label{Massless}

Finally, we study a family of solutions which is example 2 of Bronnikov-Melnikov-Dehnen solution-generating algorithm in \cite{Bronnikov}. 
There is a natural limiting procedure that takes these solutions to the massless limit of the Tidal black hole. This family of spacetimes is interesting as it has a subset of solutions which admit interpretations of wormholes, just like the CFM family of solutions. It is parametrized by a dimensionless constant $C$ and a length parameter $h$ which in the $C=1$ limit can be interpreted as an imaginary charge of the massless Reissner-Nordstr$\ddot{\text{o}}$m. 
\be
 f(r) = 1 - \frac{h^2}{r^2}, \,\, g^{-1} (r) = f(r) \left( 
1+ h\frac{C-1}{\sqrt{2r^2 - h^2}}
\right), h>0.
\ee
Metric singularity arises at the following radii 
\be
r_h = \begin{cases}
h, C \geq 0 (\text{horizon}) \\
\sqrt{\frac{1}{2} (h^2 + h^2(1-C)^2)}, C<0 (\text{wormhole throat})
\end{cases}
\ee
In particular, we note that $C=1, f=g^{-1}$ is the Reissner-Nordstr$\ddot{\text{o}}$m metric with zero mass and imaginary charge
or the massless
limit $(m=0)$ of the Tidal black hole. 
We find the asymptotic series solutions
\bea
S_1 (u) &=& u^3 \left( 1  -\frac{13 (C-1)}{8 \sqrt{2}} u + \frac{1}{560} \left(533 C^2-1066 C+1093\right) u^2 + \ldots \right) \cr
S_2 (u) &=& \frac{\sqrt{2}}{10} (1-C) u^3 y_1 (u) \log (u) + 1 + u^2 + \frac{1}{960} \left(-91 C^2+182 C+869\right) u^4 + \ldots \cr
S_3 (u) &=& \frac{11 (C-1)^3}{160 \sqrt{2}} y_1 (u) \log^2 (u) \cr
&&\qquad -\frac{11 (C-1)^2}{16} \log (u) \left( 1 + u^2  -\frac{1375 C^4-5500 C^3+46912 C^2-82824 C+68557}{13200 \sqrt{2} (C-1)} u^3 + 
\ldots \right) \cr
&&\qquad  \frac{1}{u^2} \left( 1 + \frac{11 (C-1)}{4 \sqrt{2}} u -\frac{2}{3}u + \ldots \right)
\eea
We find that the various series solutions reduce to those 
of the massless Tidal black hole
 upon setting $C=1$.  We find that with $q=-h^2$,
\bea
&&\lim_{C\rightarrow 1} S_1 (u) = y_d(u,m=0), \cr
&&\lim_{C\rightarrow 1}S_2 (u) =  y_p(u,m=0) / \alpha, \cr
&&\lim_{C\rightarrow 1} S_3 (u) =  y_r(u,m=0), 
\eea
where for $S_3 (u)$ we have set $d_2 = - 2h^2/3 = 2q/3$ and $d_5=0$. 
Taking into account the background metric, we find (see Figure \ref{fmassless} below. )
\be
\lambda =h^5  \frac{(C-1) \left(55 C^2-110 C-1\right)}{576 \sqrt{2}}.
\ee
The vanishing value $\lambda$ corresponds to the massless Reissner-Nordstr$\ddot{\text{o}}$m black hole with imaginary charge/massless Tidal black hole. For negative $C$, we have a wormhole solution. For $C \in[0,1]$, 
we obtain Kerr-like regular black hole. For $C>1$, we have Schwarzschild-like causal structure with the singularity at some finite value of $r=h/\sqrt{2}$. 
\begin{figure}[h!]
\centering
\includegraphics[scale=0.61]{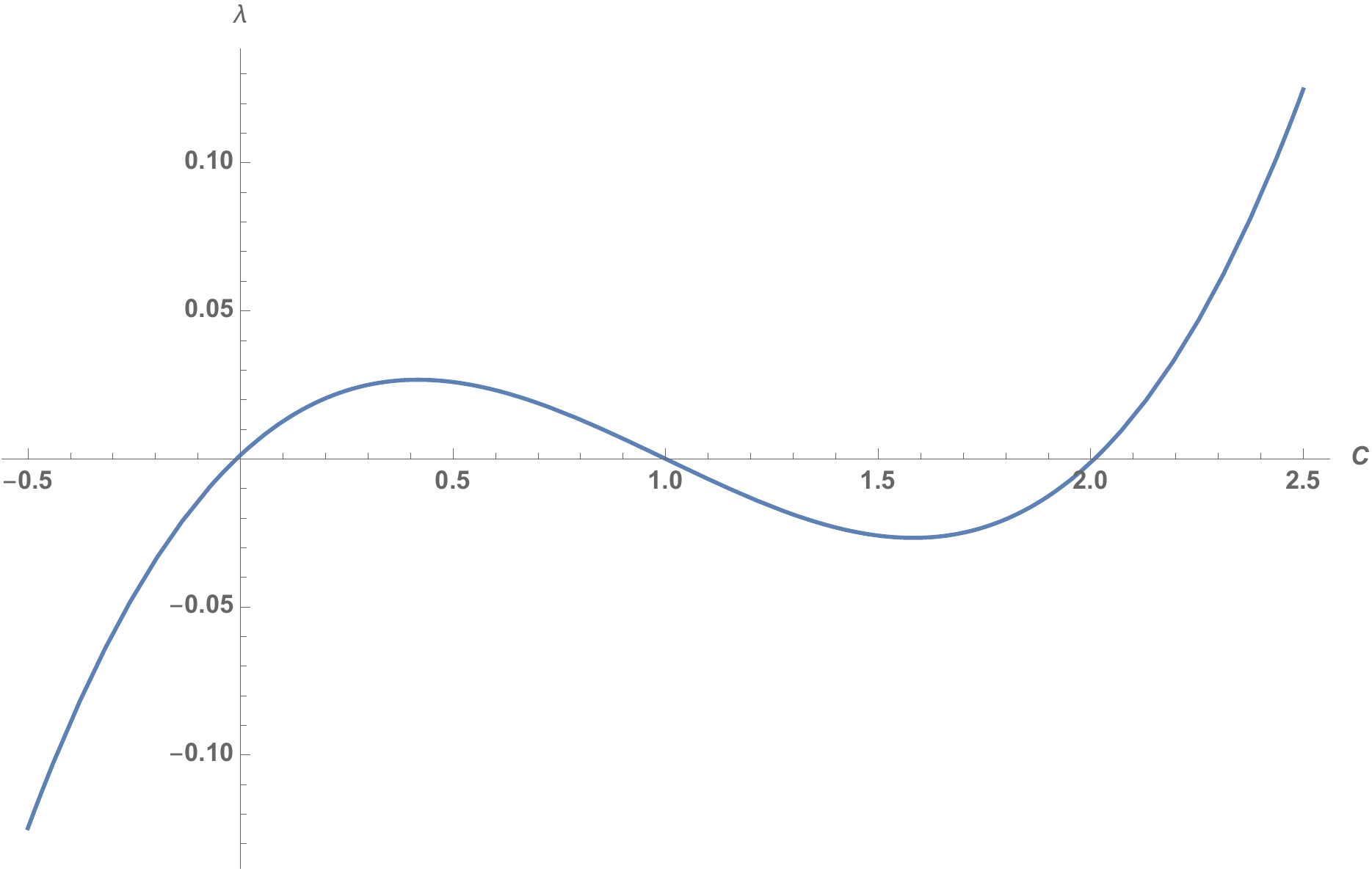}
\caption{Plot of $\lambda$ (in units of $h^5$) vs $C$. This is a cubic curve with the zero at $C=1$ corresponding to the transition case
of the massless Tidal black hole with $q=-h^2$. For $C>1$, the causal structure is identical to that of Schwarzschild. For $C<0$, we have globally regular wormholes, but $C=0$ is not a polynomial root. }
\label{fmassless}
\end{figure}

For the series expansion about the horizon, we find that $C=0$ splits the indicial roots into several branches as 
summarized in Table \ref{Table6} below. They are always positive for $C \leq 0$ and for $C>0$, take on the same set of values of
$\{-1,0,1 \}$ as in the massless Schwarzschild limit $C=1$ where we find
\be
\lim_{C\rightarrow 1} \tilde{S}_d (x) = \tilde{S}_d (x,m=0) + 2\tilde{S}_p (x,m=0)/\alpha,
\ee
and thus is independent of the near-horizon solution $\tilde{S}_r (x,m=0)$ corresponding to $S_3 (u)$. We also find that 
\be
\lim_{C\rightarrow 1} \tilde{S}_r (x) =\tilde{S}_r (x,m=0), \,\,\, 
\lim_{C\rightarrow 1} \tilde{S}_p (x) = \frac{2}{3} \tilde{S}_p (x,m=0).
\ee

\begin{table}[h!]
\begin{center}
\begin{tabular}{  |c|c| } 
 \hline
$r_h$  & Indicial Roots \\ 
\hline 
 WH, $C<0, r_{th}=\frac{1}{\sqrt{2}}\sqrt{h^2+h^2 (1-C)^2}$ &  $\{2,1,0 \}$\ \\ 
Kerr-WH, $C\in (0,1) , r_{h}=h$ &   $\{-1,0,1 \}$\\ 
Sch-BH, $C \geq 1 , r_{h}=h$ & $\{-1,0,1 \}$\\
RN-WH $C=0 , r_h=h$ &  $\{0,1,5 \}$\\
\hline
\end{tabular}
\caption{At $C=1$, we recover the massless limit of the Tidal black hole. $C=0$ marks the point where the indicial roots separate into branches. All sets of roots are non-negative or equal to that of the Tidal black hole. }
\label{Table6}
\end{center}
\end{table}

\section{Concluding remarks}
\label{Concl}

For each of the five classes of static and spherically symmetric braneworld solutions examined in this work, we computed
the (quadrupolar) Tidal Love Number, each being some rational function of the parameter that characterizes the family 
of solutions. They are generically non-vanishing and we have derived them essentially by performing an
asymptotic series expansion about the radial infinity.
The gravitational perturbation equations were shown to reduce to a single third-order homogeneous
ODE and we found that the indicial equation associated with the regular singular point at infinity 
has a universal set of roots $\{3,0,-2\}$ across all the braneworld solutions, with the tidal-deformed geometry
described by the series solution with the negative root. 

Among the braneworld solutions, there is a distinguished case in which the effective density fluctuation
can be solved analytically and the perturbation equation reduces to an inhomogeneous second-order ODE
of which indicial roots associated with the asymptotic expansion are $\{ 3,-2 \}$. This corresponds
to the Tidal black hole which is one of the more popularly studied black hole solutions in Randall-Sundrum theory 
and its TLN reads simply as $\lambda = \frac{2mq^2}{3}$ where $q$ is the tidal charge. 
We have completed the TLN calculation assuming that the 
asymptotic series solution associated with the negative indicial root $y_r (u,q))$ is regular at the horizon. This assumption
is valid if the near-horizon limit commutes with the limit of vanishing tidal charge which appears to be physically
reasonable. We also found that at linear-order in the tidal charge, 
$y_r (u,q)$ has finite number of terms and vanishes at the horizon. It would be an important future work 
to study this assumption more carefully, possibly by examining the higher-order terms in the tidal charge expansion
or numerically integrating
the ODE that governs the perturbation.

For other braneworld solutions that we have studied here, each can be parametrically connected to either the Schwarzschild 
or other limits of the Tidal black hole which thus serves as an anchor point. Once we established the case for Tidal black holes, we performed various limits (specifically, (i)$q=0$, (ii)$m=0$ and (iii)$q=m^2$ ) to relate the series solutions in each family of solutions to those for Tidal black holes and identify the one relevant
for picking up the TLN. As a necessary consistency check, we also constructed series solutions by expanding about the metric singularity
in each family of solutions,
which can be interpreted as either horizon or throat surfaces. Unlike those in the
asymptotic expansion, the indicial roots for the near-horizon expansion 
separate into branches at certains points in the parameter space, and do not 
simply take on a universal set of values. In certain parameter domains, they are all positive and thus all asymptotic 
series solutions are regular at horizon. Our method also applies when they are equal to the set $\{ -1,0,1 \}$ associated with 
the Tidal black hole, and their near-horizon series solutions can be smoothly connected to those of the Tidal black hole. 

It would be interesting to study the other cases in the parameter space where they cannot be smoothly 
deformed to limits of the Tidal black hole. This would require other approaches of analysis beyond the limiting 
procedures that we had used in this work. It is notable that in almost all cases, the asymptotic 
series solutions themselves exhibit no singularity\footnote{The only exception is the naked singularity limit of the family of 
Bronnikov-Kim wormholes.}  and it is possible that the expressions for the TLN extend to these regions of the 
parameter space. Another supporting point is that in all cases, we find at least one positive root. But we will need other methods
to study if this corresponds to the near-horizon description of the series solution $S_3 (u)$.

Although we have focussed on the computation of the quadrupolar $l=2$ TLN for various braneworld solutions, this work also contains several
results with regards to the general $l$ case as described by \eqref{ThirdOrderODE} and \eqref{DecoupledODE}. For the near-horizon expansion, 
we found that the indicial roots are almost exclusively independent of $l$, the exceptions being those with the causal structure of extremal Reissner-Nordstr$\ddot{\text{o}}$m. For the asymptotic expansion, the set of indicial roots $\{3,0,-2 \}$ generalizes to $\{ l+1,0,-l \}$
whereas for the Tidal black hole, $\{3,-2 \}$ generalizes to $\{l+1,-l \}$. The methodology to compute the higher TLNs is similar - 
the relevant series solution is the one associated with the negative indicial root and we can adopt the identical limiting procedure 
to check the consistency of this approach via the parametric connection of the various braneworld solutions to the Tidal black hole. 
From the phenomenological point of view, the higher TLNs may not be as important since they should remain beyond the detectors' precision reach. But our study of the indicial roots in the near-horizon expansion suggests that certain aspects of the indicial equation may contain physical data related to global causal structures beyond horizon geometry. Thus, it is nevertheless an interesting conjecture to explore
as it may yield further insights into the nature of gravitational degrees of freedom in braneworld black hole solutions.

A natural future work is to study the implication of the logarithmic terms and other terms which lie between $r^2$ and $\frac{1}{r^3}$  in the tidal-deformed metric, as collected in the Appendix \ref{AppB}. A number of them have natural interpretations. For example, in $h_{tt}$, terms of the
order $\frac{1}{r}, \frac{1}{r^2}$ can be absorbed in renormalizing the ADM mass and tidal charge (the TLN expressions remain valid as functions of the bare parameters), whereas the constant term can be absorbed in an overall rescaling of the metric component without changing the TLN expressions. It was noted in \cite{Thorne2} that the logarithmic terms may in principle
be present generically but were not present for the Kerr solution in ordinary perturbative Einstein gravity. Also, it was argued that the $r$ term can be gauged away. We have computed only the
traditionally defined TLNs but a more careful study may reveal the signature of these other terms in gravitational waves emitted in binary systems of these objects comparable to that of the TLN. It would be important to study these effects together with the TLN in characterizing gravitational waveforms for example.

With regards to phenomenology, it was suggested in \cite{Cardoso} that LIGO's precision could potentially bound the quadrupolar TLN $k_2$ to be of the order of ten or less, with the Einstein Telescope possibly improving it by a factor of a hundred, and more with LISA's capabilities. The discovery of a non-vanishing $k_2$ in the GW waveform of merger event presumed to be that of a black hole binary would provide exciting grounds for further study of whether the deviation is a signature of a braneworld among other possibilities. It would be interesting to generalize our work to include realistic braneworld solutions arising from gravitational collapse and in the broader context of warped compactifications in string theory beyond the specific Randall-Sundrum model we considered.

\section*{Acknowledgments}
I am very grateful to Ori Ganor, Petr Horava, A.T.Phan and Neal Snyderman for their moral support over the years, 
and acknowledge a research fellowship at the School of Physical and Mathematical Sciences, NTU during the course of completion of this work. I also thank an anonymous referee for various suggestions and advice, especially on the point of our assumption of regularity at the horizon of the asymptotic series solutions. 
\newpage

\appendix

\section{On series solutions for the metric perturbation}
\label{AppA}
In this Appendix, we collect some explicit terms in the near-horizon and asymptotic series expansions for the metric perturbation
of the Tidal black hole and geometries.

\subsection{Near-horizon expansion of the Tidal black hole}
\label{AppA1}
For notational convenience,
we express $q$ in units of mass parameter $m$ below (to restore it, simply take $q \rightarrow q/m^2$). 
We define $x=(r-r_h)/m$, $r_h$ being the horizon of the Tidal black hole. 
The two homogeneous solutions are thus of the form
\be
\tilde{y}_r (x,q) = x \left( 1 + a_1 x + a_2 x^2  + \ldots \right), \,\,\,
\tilde{y}_d (x,q) = \mathcal{N} \tilde{y}_r (x,q) \log (x) + \frac{1}{x} \left( 1 + b_1 x + b_2 x^2 + b_3 x^3 + \ldots \right),
\ee
where
\bea
a_1 &=&\frac{\sqrt{1-q} q+7 q-12 \sqrt{1-q}-12}{6 \left(-q^2+3 \sqrt{1-q} q+5 q-4 \sqrt{1-q}-4\right)}, \cr
a_2 &=&-\frac{q \left(\left(3 \sqrt{1-q}+11\right) q^2-4 \left(5 \sqrt{1-q}+7\right) q+16 \left(\sqrt{1-q}+1\right)\right)}{12 (q-1) \left(\left(\sqrt{1-q}+7\right) q^3-8 \left(3 \sqrt{1-q}+7\right) q^2+16 \left(5 \sqrt{1-q}+7\right) q-64 \left(\sqrt{1-q}+1\right)\right)}, \cr
b_1 &=&\frac{-3 \sqrt{1-q} q-13 q+20 \sqrt{1-q}+20}{2 \left(-q^2+3 \sqrt{1-q} q+5 q-4 \sqrt{1-q}-4\right)}, \,\,
b_2=0,\cr
\mathcal{N} &=& -\frac{6 \left(-q^3+6 \left(\sqrt{1-q}+3\right) q^2-16 \left(2 \sqrt{1-q}+3\right) q+32 \left(\sqrt{1-q}+1\right)\right)}{(q-1) \left(\left(\sqrt{1-q}+7\right) q^3-8 \left(3 \sqrt{1-q}+7\right) q^2+16 \left(5 \sqrt{1-q}+7\right) q-64 \left(\sqrt{1-q}+1\right)\right)}, \nonumber
\eea
And the particular solution reads
\be
\tilde{y}_p (x,q) = C_0 + \left(C_1 + \mathcal{K}_1 \log (x) \right) x + \left(C_2 + \mathcal{K}_2 \log (x) \right) x^2 + \ldots,
\ee
where
\bea
C_0 &=& - \frac{3 \alpha \left(-q+2 \sqrt{1-q}+2\right)}{2 (q-1)}, \,\,\, C_1=0,\cr
K_1 &=& \frac{3 \alpha  \left(-q^2+4 \sqrt{1-q} q+8 q-8 \sqrt{1-q}-8\right)}{(q-1) \left(q^2-3 \sqrt{1-q} q-5 q+4 \sqrt{1-q}+4\right)},\,\,\, ,\cr
K_2&=&-\frac{\alpha  \left(q^4-12 \left(\sqrt{1-q}+5\right) q^3+16 \left(11 \sqrt{1-q}+24\right) q^2-64 \left(8 \sqrt{1-q}+11\right) q+384 \left(\sqrt{1-q}+1\right)\right)}{2 (q-1)^2 \left(\left(\sqrt{1-q}+7\right) q^3-8 \left(3 \sqrt{1-q}+7\right) q^2+16 \left(5 \sqrt{1-q}+7\right) q-64 \left(\sqrt{1-q}+1\right)\right)} \cr
C_2&=&
\alpha  \Bigg(-6 q^8+\left(68 \sqrt{1-q}+341\right) q^7+\left(238-798 \sqrt{1-q}\right) q^6-224 \left(50 \sqrt{1-q}+229\right) q^5\cr
&&+192 \left(787 \sqrt{1-q}+1903\right) q^4-256 \left(2400 \sqrt{1-q}+4211\right) q^3+512 \left(2209 \sqrt{1-q}+3083\right) q^2\cr
&&-4096 \left(238 \sqrt{1-q}+277\right) q+319488 \left(\sqrt{1-q}+1\right)\Bigg)\cr
&& \times
\Bigg[ 6 (q-1)^2 \Bigg(q^7-14 \left(\sqrt{1-q}+7\right) q^6+224 \left(2 \sqrt{1-q}+7\right) q^5-1344 \left(3 \sqrt{1-q}+7\right) q^4\cr
&&+3840 \left(4 \sqrt{1-q}+7\right) q^3-5632 \left(5 \sqrt{1-q}+7\right) q^2+4096 \left(6 \sqrt{1-q}+7\right) q-8192 \left(\sqrt{1-q}+1\right)\Bigg) \Bigg]^{-1} \nonumber
\eea

\subsection{Near-horizon series expansions for CFM black holes ($\beta < 5/4$) and Massless geometries ($C>0$)}
 \label{AppA2}

Corresponding to the indicial roots $\{1,0,-1\}$, we have the ansatz
\bea
\tilde{S}_r (x)  &=& x ( 1 + a_1 x + a_2 x^2 + a_3 x^3 + \ldots )，  \cr
\tilde{S}_p (x) &=& V_1 \tilde{S}_r (x) \log (x) + (1 + b_1 x + b_2 x^2 + \ldots ), \cr
\tilde{S}_d (x) &=& \frac{1}{x} \left( 1 + d_1 x + d_2 x^2 + \ldots \right) + 
W_2 \log (x) \left( c_0 + c_1 x + c_2 x^2 + \ldots \right) + W_3  \tilde{S}_r (x) (\log (x))^2 \nonumber
\eea
For the massless geometries we have
\bea
a_1 &=& \frac{2-C}{6C}, \,\,
a_2 = \frac{-1+6C-4C^2}{4C^2}, \,\,
a_3= \frac{34-161C - 200C^2+ 296 C^3}{120C^3}, \cr
V_1 &=& -\frac{2(C-1)}{C}, \,\, b_1= 0, b_2 = -\frac{2(19-33C+11C^2)}{9C^2}, b_3= \frac{67-139C+6C^2+54C^3}{24C^3} \cr
c_0&=& W_2 = W_3= 0, \,\, d_1=\frac{3}{2}, \,\, d_3 = \frac{23C-25}{12C}.
\eea
For the CFM black holes, we have 
\bea
a_1 &=& \frac{4\beta -7}{6(4\beta-5)}, \,\,
a_2 = \frac{-37+69\beta -32\beta^2}{3(4\beta-5)^2}, \,\, a_3 = \frac{-2511+6307\beta-5204\beta^2+1408\beta^3}{30(4\beta-5)^3}, \cr
V_1 &=& \frac{3-4\beta}{4\beta-5}, \,\, b_1=0, \,\, b_2= \frac{-327+880\beta-592\beta^2}{36(-5+4\beta)^2}, \,\,
b_3 = \frac{6592\beta^3 - 27552 \beta^2 + 35192\beta - 14217}{144 (4\beta-5)^3}, \cr
W_2 &=& W_3 = 0, d_1 = \frac{1}{2}, \,\, d_2 = 0, \,\, d_3 = \frac{4(\beta-1)}{4\beta-5}
\eea

\subsection{On the metric component $h_{tt}$ for various tidal-deformed braneworld solutions}
\label{AppA3}

In this Appendix, we collect various expressions for the metric perturbation $h_{tt} = f(r)H_0 $ for $l=2$ perturbations. 
The quadrupolar TLN is read off from the $1/u^2$ and $u^3$ coefficients in each case. 
\begin{enumerate}[(I)]
\item \underline{CFM black holes}
\bea
h_{tt} &=& \frac{1}{u^2}+\frac{(3-11 \beta) }{2 u}+\left(-\frac{1}{2} \left(11 \beta^2-14 \beta+3\right)  \log (u)+(11 \beta-7) \right) 
+\frac{1}{12}  u \left(55 \beta^3-180 \beta^2+158 \beta-33\right)  \cr
&&\qquad +\frac{1}{48} u^2 \left(-(\beta-1) \right) \bigg(220 \beta^3-\left(36 \left(11 \beta^2-14 \beta+3\right)\right) \log (u)-1523 \beta^2+2358 \beta-891\bigg)\cr
&&\qquad +\frac{1}{120} u^3 \left(-(\beta-1) \right) \bigg(-1100 \beta^3+9815 \beta^2-\left(30 \left(33 \beta^3-97 \beta^2+79 \beta-15\right)\right) \log ^2(u) \cr
&&\qquad +\left(1100 \beta^4-11047 \beta^3+25706 \beta^2-23235 \beta+7191\right) \log (u)-16790 \beta+5775\bigg)+O\left(u^4\right).
\nonumber \\
\eea

\item \underline{$\gamma$-wormholes}
\bea
h_{tt} &=& 
\frac{2}{u^2}+\frac{5-11 \gamma }{u}+\bigg(10 \gamma +\bigg(-11 \gamma ^2+14 \gamma -3\bigg) \log (u)-7\bigg) \cr
&&+\frac{1}{6} u \left(55 \gamma ^3-169 \gamma ^2+141 \gamma -6 \left(11 \gamma ^2-14 \gamma +3\right) \log (u)-15\right) \cr
&&-\frac{1}{24} u^2 \left((\gamma -1) \left(220 \gamma ^3-1453 \gamma ^2+2106 \gamma -24 \left(11 \gamma ^2-36 \gamma +9\right) \log (u)-639\right)\right) \cr
&&-\frac{1}{600}\bigg( (\gamma -1 ) \bigg(-25 \left(233 \gamma ^3-1284 \gamma ^2+2232 \gamma -711\right)+2\bigg(12327-56665 \gamma +93611 \gamma ^2 \cr
&&+5500 \gamma ^4-45683 \gamma ^3\bigg)\log (u)-60 \left(121 \gamma ^3-407 \gamma ^2+355 \gamma -69\right) \log ^2(u)\bigg)\bigg) u^3 + O\left(u^4\right). \nonumber \\
\eea

\item \underline{Bronnikov-Kim wormholes}
\bea
h _{tt} &=&\frac{1}{u^2} +\frac{11 R^2-12 R+12}{(4-4 R) u} \cr
&&\,\, +\frac{1}{24(R-1)^2}8 \left(33 R^3-97 R^2+128 R-64\right)-3 (R-2)^2 \left(11 R^2-12 R+12\right) \log (u) \cr
&&\,\, +\frac{\left(55 R^6-720 R^5+2304 R^4-3552 R^3+2736 R^2-1152 R+384\right) u}{96 (R-1)^3} \cr
&&\,\, +\frac{1}{192 (R-1)^4}\bigg( -24000 + 96000 R - 176592 R^2 + 193776 R^3 - 134772 R^4 + 58584 R^5 - 
 14739 R^6 \cr
&&\,\, +1743 R^7 - 55 R^8 + 
   36 (-2 + R)^4 (-12 + 24 R - 23 R^2 + 11 R^3) Log[u] \bigg) u^2 \cr
&&\,\,+\frac{1}{4800 (R-1)^5}\Bigg[100 \bigg(55 R^9-2128 R^8+21462 R^7-102453 R^6+284012 R^5-499588 R^4\cr
&&\,\,+577088 R^3-431792 R^2+191680 R-38336\bigg) +\bigg(-1375 R^{10}+60735 R^9-718471 R^8\cr
&&\,\,+4174672 R^7-14597128 R^6+33515840 R^5-52485312 R^4+56357888 R^3-40272512 R^2 \cr
&&\,\,+17455360 R-3491072\bigg) \log (u) \cr
&&\,\, +30 (R-2)^4 \left(165 R^5-1357 R^4+3488 R^3-4504 R^2+3312 R-1104\right) \log ^2(u)\Bigg] u^3+O\left(u^4\right).
\eea

\item \underline{Massless geometries}
\bea
h_{tt} &=&
\frac{1}{u^2}+\frac{11 (C-1)}{\left(4 \sqrt{2}\right) u}+\frac{1}{48} \left(-\left(33 (C-1)^2\right) \log (u)-80\right)+\frac{\left(-55 C^3+165 C^2-637 C+527\right) u}{192 \sqrt{2}} \cr
&&\,\, -\frac{1}{768}
\left((C-1)^2 \left(55 C^2-110 C+1319\right)\right) 
u^2 \cr
&&\,\, +\frac{1}{19200 \sqrt{2}}
(C-1) u^3 \bigg(100 \left(55 C^2-110 C-1\right)+\bigg(1375 C^4-5500 C^3+46912 C^2 \cr
&&\,\,-82824 C+68557\bigg) \log (u)+\left(1320 (C-1)^2\right) \log ^2(u)\bigg)+O\left(u^4\right)
\eea

\end{enumerate}

\section{More about indicial roots}
\label{AppB}

 In the following, we compute and collect the indicial roots of all the braneworld geometries that we examine in this work by further studying \eqref{ThirdOrderODE}. The near-horizon expansion in the decoupled case (Tidal black hole) have been studied separately in Section~\ref{OnTBH}, where we found the indicial equation of the form \eqref{IndicialHor}
\be
\mathcal{R} (\mathcal{R} - 1) (\mathcal{R} -2) + \mathcal{R} (\mathcal{R} -1) \mathcal{P}_{10} + \mathcal{R} \mathcal{P}_{20}
+ \mathcal{P}_{30} = 0. 
\ee
In Table \ref{Table7} below, we present their values for all the spacetime geometries considered in our paper. 
The roots of the indicial equation are, apart from a couple of cases, independent of $l$ and for all solutions, 
there is at least one positive non-negative root which is a necessary condition for the existence of series solutions
regular at the metric singularity associated with either a horizon/throat. For cases where all 
the roots are positive or if the set of roots is $\{-1,0,1\}$, we could identify the asymptotic series solution
relevant for the TLN by taking appropriate limits. For other parameter domains, we cannot compute the TLN based rigorously
on our method. Nevertheless, since there are no singularities in the asymptotic series solutions and there is at least one positive root in all cases, it is certainly possible that the various expressions for the TLN extend to some or all of these regions. 
Other methods of analysis beyond what we have used in this work have to be explored for this purpose.

\begin{table}[h!]
\begin{center}
\begin{tabular}{ | m{3.6cm} | m{6.8cm} | m{3.1cm} | m{2.5cm} | } 
 \hline
 $\,$ & $r_h$ & $\{ \mathcal{P}_{10}, \mathcal{P}_{20}, \mathcal{P}_{30} \}$ & Indicial Roots \\ 
\hline 
\hline
 Bronnikov-Kim wormholes ($ L=\tilde{m} $) & [NS] $r_0 <\tilde{m}, r_s = 2\tilde{m} $ & $\{3,-3, 0 \}$ & $\{-2,0,2\}$\\ 
$\,$                                         & [WH] $r_0 \in [\tilde{m}, 2\tilde{m}) , r_{th} = r_1 $ & $\{\frac{3}{2},0, 0 \}$ & $\{0,\frac{1}{2},1\}$\\ 
$\,$                                         & [WH] $r_0 > 2\tilde{m} , r_{th} = r_0 $ & $\{\frac{3}{2},0, 0 \}$ & $\{0,\frac{1}{2},1\}$\\ 
$\,$                                         & [RN] $r_0 =  2\tilde{m} , r_{h} = r_0  $ & $\{6,4-l-l^2, -2(2+l+l^2) \}$ & $\{ -2,
-\frac{1}{2}(1 \pm \sqrt{9+4l+4l^2} )
\}$\\ 
\hline
Massless geometries $(L=h)$ & [WH] $C<0, r_{th}=\frac{1}{\sqrt{2}}\sqrt{h^2+h^2 (1-C)^2}$ & $\{  0,0,0 \}$ &  $\{2,1,0 \}$\\
$\,$      & [Kerr-WH] $C\in (0,1) , r_{h}=h$ & $\{  3,0,0 \}$ &  $\{-1,0,1 \}$\\
$\,$      & [Sch-BH] $C \geq 1 , r_{h}=h$ & $\{  3,0,0 \}$ &  $\{-1,0,1 \}$\\
$\,$      & [RN-WH] $C=0 , r_h=h$ & $\{  -3,0,0 \}$ &  $\{0,1,5 \}$\\
\hline
$\gamma$-wormholes \qquad $(L=m)$ & [NS] $\gamma <1, r_s = \frac{2m}{2-\gamma}$ & $\{ 0,0,0  \}$ &  $\{2,1,0\}$\\ 
$\,$ & [Sch-BH] $\gamma =1, r_h = 2m$ & $\{ 3,0,0  \}$ &  $\{-1,0,1\}$\\ 
$\,$ & [WH] $\gamma > 1, r_{th} = 2m\gamma $ & $\{ \frac{3}{2},0,0  \}$ &  $\{0,\frac{1}{2},1\}$\\ 
\hline
CFM black holes $(L=m)$ & [Sch-BH] $\beta <1, r_h = 2m$ & $\{  3,0,0  \}$ &  $\{-1,0,1\}$\\ 
$\,$ & [Kerr-WH] $\beta \in (1,\frac{5}{4} ), r_h = 2m$ & $\{  3,0,0  \}$ &  $\{-1,0,1\}$\\ 
$\,$ & [RN-WH] $\beta  = \frac{5}{4}, r_h = 2m$ & $\{ \frac{9}{2}, \frac{1}{4}(10-l-l^2), \frac{1}{4}(-2-l-l^2)  \}$ &  $\{
-1,
-\frac{1}{4}(1 \pm \sqrt{9+4l+4l^2} )
\}$\\ 
$\,$ & [WH] $\beta > \frac{5}{4} , r_{th} = \frac{m}{2}(4\beta -1 )$ & $\{ \frac{3}{2},0,0  \}$ &  $\{0,\frac{1}{2},1\}$\\ 
\hline
\end{tabular}
\caption{In the above, we use the abbreviations  WH (wormholes), Sch-BH (Schwarzschild black holes), NS (Naked Singularities), RN (extremal Reissner-Nordstr$\ddot{\text{o}}$m). Kerr-WH refers to completely regular wormhole geometries of which CP diagram resembles the form of that of Kerr. RN-WH refers to spacetimes with the casual structure of extremal 
Reissner-Nordstr$\ddot{\text{o}}$m with a horizon cloaking a time-like singularity.The radius parameters $r_s, r_{th}$ refer to the singular surface of a naked singularity and the radius of a wormhole throat respectively. This table excludes the roots $( \pm 1 )$ for the Tidal black hole which are associated with a second-order ODE as was
presented in Section~\ref{OnTBH}.  }
\label{Table7}
\end{center}
\end{table}

For each family of solution characterized by some real parameter $\alpha$ (as summarized in Table \ref{Table1}, $\alpha \in \{ r_0, h, \gamma, \beta \} $)
 the coefficient functions $\{ \mathcal{P}_{1} (x, \alpha), \mathcal{P}_{2} (x,\alpha), \mathcal{P}_{3} (x,\alpha) \}$
are discontinuous as bivariate functions of $x$ and $\alpha$ at the regular singular point $x=0$ at values of $\alpha$  which mark 
transitions between different spacetime interpretations. This leads to different sets of roots characterizing different types
of spacetime geometries. Interestingly, we find that for the various solutions we study, the indicial roots tend to display some level of universality --- 
similar sets of values are associated with the solutions belonging to different families but sharing identical causal structures. 
From Table \ref{Table7}, one can easily recognize the following pairings between the indicial roots and spacetime interpretations. 
\begin{itemize} 
\item $\{ -1, 0, 1 \} \sim$ Schwarzschild black holes and globally regular Kerr-like wormholes
\item $\{0, \frac{1}{2} , 1\} \sim$ globally regular and traversable wormholes
\end{itemize}
The only solutions with $l$-dependent roots are the extremal Reissner-Nordstr$\ddot{\text{o}}$m solution which is a member of the Bronnikov-Kim family of solutions and a member of the CFM family of black hole solutions of which CP diagram is identical to that of extremal Reissner-Nordstr$\ddot{\text{o}}$m but which are geometrically different. A subset of solutions defined by $C\leq 0$ in the `Massless geometries' family appears to fall out of this classification. The case of $C=0$ has extremal Reissner-Nordstr$\ddot{\text{o}}$m-like causal structure and metrices with $C<0$ are globally regular and traversable wormholes. It would be interesting to study the relation between indicial roots and spacetime causal structures with more examples and in a deeper systematic classification. These roots reflect the tidal response of the object in the near-horizon regime and 
appear to capture aspects of causal structures of the various objects beyond their local horizon geometries.

\end{document}